\documentclass[10pt]{article} 
\usepackage{yfonts} 
\usepackage{epsfig} 
\usepackage{subfigure} 
\usepackage{supertabular} 
\usepackage{amsmath} 
\usepackage{amssymb} 
\usepackage{bbold} 
\usepackage{cite} 
\usepackage[sans]{dsfont} 
\usepackage{graphicx} 
\usepackage[dvips]{color} 
\usepackage{soul} 
\usepackage{multirow} 
\usepackage{lscape} 
\newcommand{\cleqn}{\setcounter{equation}{0}}
\newcommand{\newc}{\newcommand}
\newc{\om}{\omega}
\newc{\im}{{\bf i}}
\newc{\sig}{\sigma}
\newc{\epi}{\epsilon}
\newc{\beq}{\begin{equation}}
\newc{\eeq}{\end{equation}}
\newc{\beqn}{\begin{eqnarray}}
\newc{\eeqn}{\end{eqnarray}}
\newc{\bsym}{\boldsymbol}
\newc{\ol}{\overline}
\newc{\unl}{\underline}
\newc{\ord}{\mathcal{O}}
 \def\bvec#1{\raise1.5ex\hbox{$\rightarrow$}\mkern-16.5mu #1}

\begin{document}

\title{\hfill ~\\[-30mm]
       \hfill\mbox{\small UFIFT-HEP-xx-xx}\\[30mm]
              \textbf{Flavor $\bsym{\Delta(54)}$ in $SU(5)$ SUSY Model.}}
\date{\today}
\author{\\
        J. A. Escobar\footnote{E-mail: {\tt jescobar@phys.ufl.edu}}~~
         \\ \\
  \emph{\small{}Institute for Fundamental Theory, Department of Physics,}\\
  \emph{\small University of Florida, Gainesville, FL 32611, USA}}
\maketitle

\begin{abstract}
\noindent We design a supersymmetric $SU(5)$ GUT model using
$\Delta(54)$, a finite non-abelian subgroup of $SU(3)_f$. Heavy right
handed neutrinos are introduced which transform as three-dimensional 
representation of our chosen family group. The model 
successfully reproduces the mass hierarchical mass structures of the 
Standard Model, and the CKM mixing matrix. It then provides predictions 
for the light neutrino with a normal hierarchy and masses such
that $m_{\nu,1}\approx 5\times10^{-3}~ eV$,  
$m_{\nu,2}\approx 1\times10^{-2}~ eV$,
and $m_{\nu,3}\approx 5\times10^{-2}~eV$. We also provide predictions for 
masses of the heavy neutrinos, and corrections to the tri-bimaximal 
matrix that fit within experimental limits, e.g. a reactor angle of 
$-7.31^{o}$. A simple modification to our model is introduced at the 
end and is shown to also produce predictions that fall well within those 
limits.  
\end{abstract}

\clearpage 

\section{Introduction.}
\cleqn
The origins of the mass structure in the Standard Model (SM)
is currently without explanation. However, current neutrino oscillation 
data provides clues that a finite non-abelian symmetry may be 
responsible. The oscillation evidence that we speak of comes in 
the form of the lepton mixing matrix ($\mathcal{U}_{{mnsp}}$), which 
plays the same role as the CKM matrix for the quarks. The most
promising and current theoretical fit of $\mathcal{U}_{{mnsp}}$ is 
called the tri-bimaximal lepton mixing matrix
($\mathcal{U}_{\text{tri-bi}}$) \cite{tribimax}, 
and it is in this form one is clearly lead to the possibility of a 
non-abelian finite group being the key to solving what is
often referred to as the Flavor Problem.  

In this paper we postulate a finite subgroup of an
$SU(3)_f$ family group: $\Delta(54)$ is responsible for the masses 
and mixing data observed \footnote{In the literature one can find
many possible finite groups as the origins of the large
mixing angles in the lepton sector, e.g. for $A_4$
\cite{a4papers}, $S_4$ \cite{s4papers}, $\Delta(27)$
\cite{D27papers}, and $PSL_{2}(7)$ \cite{psl27papers}.}. We 
reach the goal of creating a model by way of the Froggatt-Nielsen (FN) 
formalism which is an effective field theory suppressed by some mass 
scale \cite{FNmech}. The mass scale of the FN formalism allows 
for the introduction of a single parameter that controls the 
perturbative nature of the theory. The model is ambitious in that we 
try to only use this one parameter throughout. Now, these details and 
more of the model building process are laid out in several 
sections, which we now summarize.

In order that we produce experimentally viable results it is
essential to keep in mind all experimental data and constraints. To 
this end, Section \ref{sect:phenoconst} serves a dual role as a 
summary of the data to be reproduced and a discussion on how it
should be accomplished.  For the sake of organization the
section splits the phenomenology into quark and
lepton sectors. We list in each what it is we want to reproduce 
from experimental results and how it can be done.

With constraints at hand, in Section \ref{sct:modelbuild} we give a 
closer look at $\Delta(54)$ and determine how it may be
implemented. We also discuss how and why we split the matter content into
specific representations of our flavor group. Then under these choices
we make use of a toy model to demonstrate how we satisfy the 
constraints found in the previous section.

The fourth section contains the model which as a final result 
can be summarized as coming from $SU(5)\otimes \Delta(54) 
\otimes Z^u_3 \otimes Z^d_2\otimes Z_2$. The underlying
assumption of our model is that we have supersymmetry at this
scale. So that we show the super-potential for our theory and take a 
look at the contributions to each sector. At the end we provide
the predictions to the neutrinos and the angles of the lepton
mixing matrix. Specifically we find a normal hierarchy
structure with neutrino masses being: $m_{\nu,1}\approx 
5\times10^{-3}~ eV$,  $m_{\nu,2}\approx 1\times10^{-2}~ eV$,
and $m_{\nu,3}\approx 5\times10^{-2}~eV$. As for the angles we
find the reactor angle of $\theta_{13} \approx -7.31^{o}$,
with the post-dicted solar angle of $\theta_{\odot} \approx
34.36^{0}$, and an atmospheric angle of $\theta_{atm}\approx-45.15^{o}$.

The final section includes a simple modification to the
model discussed above. We explore the alteration 
and show that it too may provide a viable model by taking a
specific example and listing its predictions to the angles of
the lepton mixing matrix.

\section{Phenomenological Constraints.}
\label{sect:phenoconst}
\cleqn
The goal is to produce phenomenologically correct Yukawa
matrices for the quark sector and at the same time produce
viable neutrino masses and to first order the
tri-bimaximal lepton mixing matrix. As for the charged leptons,
the choice of an $SU(5)$ GUT will automatically produce a
Yukawa from the down-quark sector. The focus of this section is 
then the phenomenology involved in each matter
sector and how to consolidate the data into mass matrices.

\subsection{Quark sector.}\label{ssect:QuarkSector.}
Current experiments allows for only two but important
pieces of data. These come in the form of the approximate
masses for the quarks and the quark mixing matrix known as the
CKM matrix. 

It is well known that extrapolating mass data to the unification 
scale one can parametrize all masses in terms of the Cabibbo angle
$\lambda_c\approx.226$, producing the hierarchical structure

\begin{equation}\label{eq:leptonmass}
\dfrac{m_{e}}{m_{\tau}} \approx \mathcal O(\lambda_c^{4,5}),
~~~~~~\dfrac{m_{\mu}}{m_{\tau}} \approx  \mathcal O(\lambda_c^2),
~~~~~~ m_{\tau} \sim m_{b},
\end{equation}

\begin{equation}\label{eq:intrafam}
\dfrac{m_{d}}{m_{b}} \approx \mathcal O(\lambda_c^{4,5}),
~~~~~~\dfrac{m_{s}}{m_{b}} \approx  \mathcal O(\lambda_c^2), 
~~~~~~\dfrac{m_{b}}{m_{t}}\approx \mathcal O(\lambda_c^3),
\end{equation}

\begin{equation}\label{eq:topmasses}
\dfrac{m_{u}}{m_{t}} \approx \mathcal O(\lambda_c^8),
~~~~~~\dfrac{m_{c}}{m_{t}} \approx  \mathcal O(\lambda_c^4), 
\end{equation}
Included above is the relation between mass of the tau lepton and 
bottom quark which are approximately equal and the intra-family 
hierarchy, both the last relations in Eq.~(\ref{eq:leptonmass}) 
and Eq.~(\ref{eq:intrafam}) respectively. 

The choice of an $SU(5)$ model will guarantee 
the lepton masses and down-type quark masses are in fact related and
so ensure that the mass of the tau and bottom quark are
identical. So that with an $SU(5)$ model we will try to 
reproduce, in the form of eigenvalues of two Yukawa matrices,
all the information found in Eq.~(\ref{eq:leptonmass}) - (\ref{eq:topmasses}). 

The last experimental piece of data at our disposal is the CKM 
matrix. It is a mixing matrix composed roughly out of
differences in angles that occurs from diagonalizing the Yukawa
matrices of both quark sectors. The information contained there 
to third order approximation is 

\begin{equation}
\mathcal U_{ckm} 
\approx
\ord
\begin{pmatrix}
1              &   \lambda_c       &   \lambda_c^3 \\
-\lambda_c     &   1               &   A \lambda_c^2 \\
 \lambda_c^3  &   -A\lambda_c^2   &  1 \\
\end{pmatrix}
\ ,
\end{equation}
$A$ is the appropriate parameter found in the Wolfenstein 
prescription. Because of the very nature of its origins
there is a limit in how much information we can derive about 
the structure of the quark Yukawas. Nevertheless, there are clues as 
to the texture structures and if we add to these the necessary
eigenvalues required we can limit the possible choices
for Yukawa matrices~\cite{RRGLVS}. 

Taking all these constraints, and following guidelines found in
\cite{RRGLVS} we find that at the very minimum we would need
\begin{equation}\label{eq:oxyukawa1}
Y^{(2/3)}
\approx
\mathcal{O}
\begin{pmatrix}
0 & \lambda_c^6 & \lambda_c^{{\scriptscriptstyle\geq} 4}    \\
 \lambda_c^6 &  \lambda_c^4 &  \lambda_c^{{\scriptscriptstyle\geq} 2} \\ 
\lambda_c^{{\scriptscriptstyle\geq} 4} 
& \lambda_c^{{\scriptscriptstyle\geq} 2} & 1 \\
\end{pmatrix}
\ ,
~~~~
Y^{(-1/3)}
\approx
\mathcal{O}
\begin{pmatrix}
0 & \lambda_c^3 & \lambda_c^{ {\scriptscriptstyle\geq}3}    \\
 \lambda_c^3 & \lambda_c^2 &  \lambda_c^{ {\scriptscriptstyle\geq}2} \\ 
\lambda_c^{{\scriptscriptstyle\geq}1 }  & 
\lambda_c^{{\scriptscriptstyle\geq}0} & 1        \\
\end{pmatrix}
\ ,
\end{equation}
assuming that coefficients are of $\mathcal{O}(1)$. It should
be noted that the above is a bit misleading, at least one of the $(2,3)$
positions must be $\lambda_c^2$. Now, the model building will have to 
satisfy the hard texture constraints given above and fall within 
the limits placed.  

We do so not using the Cabbibo angle as our expansion
parameter for the whole model but instead $\delta\approx.20$. 
There is some arbitrariness to this, the only constraint being 
that $\delta>.182$, but we chosen its stated value so that the mass 
relations are consistent at energies of the GUT scale of 
$2\times 10^{16}~GeV$ and its value must remain close to the 
Cabibbo angle if we expect Eq.~(\ref{eq:oxyukawa1}) to remain true. 

\subsection{Lepton sector.}
Unification via $SU(5)$ will automatically produce information
about the charged leptons once the down-quark Yukawas are known. So 
we will only concentrate on both the neutral leptons and heavy 
neutrinos.

In terms of experimental data the lepton sector does not share
the same richness as the quark sector, but we do have available to us 
two key pieces of data\footnote{
Cosmological data also provide limits on the sum of neutrino
masses and the size of the most massive neutrino \cite{cosmo}

\begin{equation} \label{eq:comologycons}
\sum m_{\nu,i} < (.17-2.0)~ eV,~i=1,2,3 \ ,~~~~
.04<m_{\nu,\text{heaviest}} < (.07-.70)~eV
\ .
\nonumber
\end{equation} }. First, experimental results have 
given us the mass squared differences \cite{neutdata}

\begin{equation}
\Delta m_{21}^2
\approx 
7.59^{+.19}_{-.21}\times 10^{-5}~eV^2, 
~~~~~~
|\Delta m_{23}^2|
\approx 
2.43\pm.13\times 10^{-3}~eV^2
\ ,
\end{equation}
notice that second relation does not allow us to determine the
exact hierarchical structure. Nevertheless, a useful constraint 
that can be derived from the above is

\begin{equation}\label{eq:ratiosdiff}
29.56\leq \dfrac{|\Delta m_{23}^2|}{\Delta m_{21}^2} \leq 34.68
\ ,
\end{equation}
the average value being $\approx 32.02$. 

The second piece of experimental data comes in the form of the 
lepton mixing matrix $\mathcal U_{mns}$, which we shall assume 
to be approximately the tri-bimaximal matrix 

\begin{equation}
\mathcal U_{mnsp}
\approx
U_{\text{tri-bi}}
=
\begin{pmatrix}
\sqrt{\dfrac{2}{3}}  & \dfrac{1}{\sqrt{3}} & 0 \\
-\dfrac{1}{\sqrt{6}} & \dfrac{1}{\sqrt{3}} & -\dfrac{1}{\sqrt{2}} \\
-\dfrac{1}{\sqrt{6}} & \dfrac{1}{\sqrt{3}} & \dfrac{1}{\sqrt{2}} \\
\end{pmatrix}
\ .
\end{equation}

The see-saw mechanism requires the existence of the regular 
neutral lepton Yukawa matrix $Y^{(0)}$ and an invertible Majorana 
matrix  $Y_{maj}$ \cite{Seesaw}. These together are needed for the 
light neutrino mass approximation of

\begin{equation}
\mathcal Y_{\nu}
\approx
-
\dfrac{v^2}{\mathcal M}~ Y^{(0)} ~\left (Y_{maj} \right
)^{-1} ~Y^{(0) ~T}
\ ,
\end{equation}

which we diagonalize by $\mathcal U_{mns}$, i.e.

\begin{equation}
\mathcal Y_{\nu}
=
\mathcal U_{mnsp} 
~m_{\nu}~
\mathcal U_{mnsp}^T
\ .
\end{equation}
The diagonal term $m_{\nu}$ will in general contain three different
eigenvalues (masses) and after selecting these eigenvalues
we can produce the light neutrino matrix from the tri-bimaximal
matrix:

\begin{equation}
m_{\nu}
=
\begin{pmatrix}
 m_1  && \\ &m_2 & \\ & & m_3
\end{pmatrix}
~~
\Rightarrow
~~
\mathcal Y_{\nu}
=
\begin{pmatrix}
\Delta_1 & \Delta_2 & \Delta_2 \\
\Delta_2 & \Delta_3 & \Delta_1+\Delta_2-\Delta_3 \\
\Delta_2 & \Delta_1+\Delta_2-\Delta_3 & \Delta_3 \\
\end{pmatrix}
\ ,
\end{equation}
in which we have that

\begin{equation}
\Delta_1=\dfrac{1}{6}(4m_1+2m_2), ~~
\Delta_2=\dfrac{1}{6}(-2m_1+2m_2), ~~
\Delta_3=\dfrac{1}{6}(m_1+2m_2+3m_3)
\ .
\end{equation}
Thus the eigenvalues as functions of entries of $\mathcal 
Y_{\nu}$ are given as

\begin{equation}
m_1= \Delta_1-\Delta_2, ~~
m_2= \Delta_1+2\Delta_2, ~~
m_3= 2\Delta_3-\Delta_1-\Delta_2 
\ .
\end{equation}

\section{Model building with $\Delta(54)$.} \label{sct:modelbuild}
\cleqn
The focus of this section is to describe in some detail 
the strategy taken to produce our model. We begin with an
attempt to familiarize ourselves with $\Delta(54)$ by
having a quick look at its salient features. A complement to
this section, i.e. with a more mathematical description of this 
group, can be found in Appendices~\ref{Acomp}-\ref{Delta54}. 

In brief, Appendix \ref{Acomp} contains a comparison of 
the group itself to that of a similar group $\Delta(27)$, which has been
investigated as a flavor group \cite{D27papers}. While
Appendix~\ref{Delta54} contains some of the mathematical
information regarding the group $\Delta(54)$ that a reader
would want to know for this paper.

Our model makes use of a supersymmetric $SU(5)$ GUT theory.
This, of course, has a direct impact on how we build a theory under our
flavor group. Now, although for the most part the choice of GUT is 
somewhat arbitrary, an $SU(5)$ theory has a method of unifying the
charged lepton and down-type quark masses in a simple elegant way.
Our choice means that we must place matter into specific
representations under $SU(5)$ \cite{SU5}, these are:

\begin{equation}
{\ol{N}} \sim \mathbf{1}, ~~~
L,~\ol{d} \sim \mathbf{\ol{5}}, ~~~
Q,~\ol{u},~\ol{e} \sim \mathbf{10}, ~~~
\ .
\end{equation}
The $L$ and $Q$ are the $SU(2)$ weak doublets and the remainder
particles are the right handed weak singlets.

\subsection{$\Delta(54)$ as a flavor
group.}\label{sect:asflavorgrp}
\cleqn
A glance at the appendix shows that the group has both
two and three-dimensional representations. This translates into
many options for assigning representations to the matter
content. Although all options can be explored we wish to
limit them, and for an $SU(5)$ theory this can be done by
examining the mass of the top quark.

The origins of its mass is at tree-level, since its value seems 
close to that of vacuum expectation value (vev) of 
the Higgs particle. Ensuring this result
satisfactorily for the three-dimensional representation is
very difficult if not impossible to do. To see that this is indeed
the case, let's for the moment describe what would happen if we
used such a three-dimensional representation.

First, our model assumes that the top quark mass comes from the
product of two ten-dimensional representations of $SU(5)$.
Let's assume that under our flavor group the ${X}\sim \mathbf{10}$ 
transforms as any of the four three-dimensional
representations, i.e. $\mathbf{3_1,~3_2,~\ol{3}_1,~\ol{3}_2}$. Then the
interaction term responsible for mass produces no singlets but
instead, schematically,  a direct sum of three-dimensional
representations

\begin{equation}
{X} \cdot {X}
\sim
\left (\ol{\mathbf{3}}_1 \oplus \ol{\mathbf{3}}_1 \right )_{S}
\oplus
\ol{\mathbf{3}}_{2,A}
\ .
\end {equation}
The bar should be understood as the complex conjugate of
whichever $\mathbf{3}$ taken 
for ${X}$. In order to get a singlet term we must have a 
flavon $\phi$ which transforms as either a $\mathbf{3}_1$ or a
$\ol{\mathbf{3}}_1$ depending on the representation chosen 
for the $\mathbf{10}$ so that via the FN mechanism

\begin{equation}
\dfrac{g}{M}\phi 
~{X} \cdot {X}
\ ,
\end{equation}
where $g$ is a coupling constant, $M$ is the mass scale for
the mechanism, and we have suppressed the Higgs. In order to explain 
the mass of the top properly the vev of the flavon field must be the 
same order as the mass scale i.e. $\langle \phi \rangle \sim M$. 
In terms of model building this fact is difficult to explain and it 
can be difficult to control the interaction terms involving $\phi$. These
difficulties are enough to make us avoid the use the three-dimensional 
representation of $\Delta(54)$ to describe the up-quarks.

We have chosen instead to have the top quark be a singlet under the flavor 
group, i.e.  $X_3\sim 1$. While the two remaining flavors together form a
two-dimensional representations $( {X_1}, {X_2} )^T
\sim \mathbf{2}_r$, $r=1,2,3,4$. Under this scheme
we have a natural way to explain the mass of the top quark
at tree-level: ${X_3 X_3} H_u$. So we take the approach that 
both quark sectors can be written in the same fashion just 
described. Our motivation 
for the choice of $\mathbf{2}\oplus\mathbf{1}$ structure is two-fold. 

First, if we had chosen instead that the $\mathbf{\ol{5}}$ transform as
$\mathbf{3}$ under $\Delta(54)$ it would be difficult to
control the power in $\delta$ of any one entry in a Yukawa
matrix without the danger of producing that same power in
another. An issue when that same power is lower than the
power required, we refer the reader to Appendix~\ref{Delta54} to
confirm this. The second weaker reason is simply that the
Yukawas of both quark sectors are similar by
having structures which are copacetic with the use of
two-dimensional representations. Texture zero structures that
occur in both quark sectors are easily achievable and
can be understood as coming from the vevs of the two-dimensional flavon. 

We summarize our choice for the $SU(5)$ matter content under $\Delta(54)$: 

\begin{eqnarray}
(\mathbf{10}_1,~\mathbf{10}_2)^T ,~ \mathbf{10}_3
&   
\stackrel{\Delta(54)}{\sim} 
&   
\mathbf{2}_p,~\mathbf{1}, \\ \nonumber
(\mathbf{\ol{5}}_1,~\mathbf{\ol{5}}_2)^T ,~ \mathbf{\ol{5}}_3
&   
\stackrel{\Delta(54)}{\sim} 
&   
\mathbf{2}_r, ~ \mathbf{1},
~~~p,~r = \left \{1,~2,~3,~4 \right \}, 
\\ \nonumber
(\mathbf{1}_1,~\mathbf{1}_2,~\mathbf{1}_3)^T
&   
\stackrel{\Delta(54)}{\sim} 
&   
\mathbf{3}_{s} ~\text{or}~ \ol{\mathbf{3}}_{s},
~~~s = 1,~2 \ , 
\end{eqnarray}
included above is the case where $p=r$. We now investigate 
the type of Yukawa matrices we can produce based on our choice 
of representations. All the possibilities for the
up-quark and down-quark Yukawas are summarized with just two
matrices respectively

\begin{equation} \label{eq:matrices}
\left (
\begin{array}{c |c}
& \multirow{3}{*}{\rotatebox{270}{$\mathbf{2}_p\otimes
\mathbf{1} =\mathbf{2}_p$}}
\\
{\mathbf{2}_p \otimes \mathbf{2}_p} & 
\\
=
\\ 
{(\mathbf{2}_p\oplus \mathbf{1})_S\oplus \mathbf{1}_{1,A}} & 
\\
\\
\hline 
\\
{\mathbf{1} \otimes \mathbf{2}_p = \mathbf{2}_p} 
& \mathbf{1}\otimes \mathbf{1} =\mathbf{1}
\\
\end{array}
\right ) , 
~~~ \text{and}~~~ 
\left (
\begin{array}{c |c}
& \multirow{3}{*}{\rotatebox{270}{$\mathbf{2}_p\otimes \mathbf{1}=\mathbf{2}_p$}}
\\
{\mathbf{2}_p \otimes \mathbf{2}_r } & \\ 
= & \\
{\mathbf{2}_{s'}\oplus \mathbf{2}_{s''} } & 
\\
\\
\hline 
\\
{\mathbf{1} \otimes \mathbf{2}_r = \mathbf{2}_r} 
& \mathbf{1}\otimes \mathbf{1} =\mathbf{1}
\\
\end{array}
\right ) , 
\end{equation}
where $s',s''=\{1,~2,~3,~4\}$. The up Yukawa must
always necessarily be the left case. While for the down
it may be either the right case when $p \neq r \neq s' \neq s''$, 
or the left when $p=r$. 

Recall that at the end of Section~\ref{ssect:QuarkSector.} it was 
mentioned that we shall
try to reproduce the texture structure and constraints of
Eq.~(\ref{eq:oxyukawa1}). In order to show how this can be
accomplished we will make use of a toy model 
that uses two matter fields, $\chi,~\psi$, and two flavons
$\theta_1$ and $\theta_2$. The goal is to then to show how to
obtain the texture structure we seek from matrices
constructed in the fashion shown by Eq.~(\ref{eq:matrices}).

\subsection{A quark sector toy model.} \label{ssect:toymodel}
We start with notation that is used
in this toy model and throughout other sections from now on. So far 
we have decided that the representations of the matter
content will be split into $\mathbf{2}\oplus \mathbf{1}$ flavor 
representations for reasons explained in the section before. So in 
order to distinguish matter that transforms as a $\mathbf{2}$ from that
as a $\mathbf{1}$ our convention uses an underline 
for doublets and no such underline for singlets, e.g. we could
write for the left handed quark $SU(2)$ doublet 
\begin{equation}
\unl{Q}
\equiv 
\begin{pmatrix}
Q_1 \\
Q_2 
\end{pmatrix} \sim \mathbf{2}_2 \ , ~~~
Q
\equiv
Q_3 \sim \mathbf{1}  \ ,
\end{equation}
it should be understood that the subscripts are flavor
indices. As can be seen the notation will be cleaner than using
subscripts or superscripts to denote the differences in
representations. For the flavon fields the variable $\phi$ will
be used for $\mathbf{3}$, $\theta$ for $\mathbf{2}$, 
and $\sigma$ for either the $\mathbf{1_1}$ or
the $\mathbf{1}$ representations. Any subscripts found on the
flavons will aid in simply distinguishing among them.

Returning to our toy model, we shall assume that our fields
should transform as shown in Table~\ref{tb:toyflavon1}:

\begin{table}[htb]
\caption{Matter content and flavons for the toy model 
        with $p,~r,~s=\{1,~2,~3,~4\}$.}
\begin{center}
\begin{tabular}{l c c | l c c}
\hline
\hline
Matter & $SU(5)$ & $\Delta(54)$ & Flavons, $\langle vev \rangle$ & $SU(5)$ & $\Delta(54)$  
\\ 
\hline
& & & & & 
\\ 
$\unl{\psi},~\psi$ &$ \ol{\mathbf 5}$ 
& 
$
\begin{array}{l r}
\mathbf{2}_r,& \mathbf{1} 
\end{array}
$
&
$\theta_1$,~ 
$\begin{pmatrix}a & b \end{pmatrix}^{T}$& $\mathbf{1}$ & $\mathbf{2}_s$ 
\\
$\unl{\chi},~\chi$ & $\mathbf{10}$ & 
$
\begin{array}{l r}
\mathbf{2}_p, & \mathbf{1}
\end{array}
$
&
$\theta_2$,~ 
$\begin{pmatrix}c&d\end{pmatrix}^{T}$ & $\mathbf{1}$ & $\mathbf{2}_s$
\\
& & & & & 
\\
\hline
\hline
\end{tabular}
\label{tb:toyflavon1}
\end{center}
\end{table}
The second flavon will be used for the case where we want to 
show with clarity a quadratic term in flavons. For the purpose
of brevity we will look at the Yukawa term for the down-type
quarks, but when possible we will discuss the up-type quark
Yukawa as well. The reason for looking at the down Yukawa is that it 
presents the most generic possible scheme since it allows both the 
case where $p=r$ and $p\neq r$. 

A Yukawa matrix for the down quark can be built from the flavon
interacting with the terms $ \unl{\chi} ~\unl{\psi} $, $\unl{\chi}
\psi $, ${\chi} \unl{\psi} $, ${\chi} \psi $. Schematically
the structure of the mass matrix is therefore 

\begin{equation}
\begin{pmatrix}
\unl{\chi} \unl{\psi} & \unl{\chi} \psi \\
{\chi} \unl{\psi} & {\chi} \psi \\
\end{pmatrix}
\ ,
\end{equation}
following the same partitioning scheme as Eq.~(\ref{eq:matrices}). With 
the all the above in mind we shall now look at several cases
involving different choices for relationships between the 
variables $p,~r,~s$. In each case we list the possible 
results and label them, only going as high up to quadratic order 
in flavon fields. Greek letters not previously defined are just 
coupling constants, and multiple such constants in front of a term
indicate there are a number of different ways to get a flavor 
invariant. The first case where $p=r$ will be the case for our model 
and so we will spend some time pointing out those important features.

\begin{itemize}

\item{$\mathbf{p=r.}$} One should notice this is 
the first case of Eq.~(\ref{eq:matrices}). There are two
possible choices we can take for the flavon; either
$p=r=s$ or $p=r\neq s$.

\begin{itemize}
\item[$(i)$]{$p=r=s$.} The tree-level results allow for
a non-zero term in the $(3,3)$ position, useful
in the case of the top quark. However this is not the only
allowed contribution, in all the zeroth order contributions are

\begin{equation}
\ord(\theta^0):~
\alpha~\unl{\chi}~\unl{\psi} 
+
\beta~{\chi}~{\psi} 
\mapsto
\begin{pmatrix}
0 & \alpha  & 0 \\
\alpha & 0 & 0 \\
0 & 0 & \beta \\
\end{pmatrix}
\ .
\end{equation}
For a realistic model, we would not like the $2\times2$ 
locations occupied at this order. To avoid these results, we are lead to
conclude that $\unl{\chi}~\unl{\psi}$ must be charged under some
symmetry that forbids it. 

For first order contributions in flavons we have:

\begin{equation}
\ord(\theta):~
\alpha~\theta_1 ~\unl{\chi}~\unl{\psi} 
+
\beta~\theta_1~\unl{\chi}~{\psi} 
+
\beta'~\theta_1~{\chi}~\unl{\psi} 
\mapsto
\begin{pmatrix}
\alpha a & 0  & \beta b \\
0 & \alpha b & \beta a \\
\beta' b & \beta' a & 0 \\
\end{pmatrix}
\ .
\end{equation}
The reader should notice how the vevs contribute to the entries
above. A choice of $a=0$ would mean that the $(1,1)$ zero could
be protected. For the up-quarks we could instead have $a=\delta^{\geq2}$ 
and $b=0$ in order to satisfy our texture constraint while hoping that 
symmetries disallow any $2\times2$ terms.
  
A look at the Kronecker products reveals that the second-order
in flavons can produce doublets and two types of singlets.

\begin{eqnarray} \label{eq:caseprods1}
\ord(\theta^2):~
\left (\alpha,~\beta,~\gamma \right ) \theta_1~ \theta_2 ~ \unl{\chi}~\unl{\psi} 
+
\rho\theta_1~\theta_2~\unl{\chi}~{\psi} 
+
\rho'\theta_1~\theta_2~{\chi}~\unl{\psi} 
+
\sigma\theta_1~\theta_2~{\chi}~{\psi} 
\mapsto
\\ \nonumber
\begin{pmatrix}
\alpha bd & \beta bc + \gamma ad  & \rho ac \\
\beta ad + \gamma bc & \alpha ac & \rho bd \\
\rho' ac & \rho 'bd & \sigma(ad+bc) \\
\end{pmatrix}
\ .
\end{eqnarray}
The $(\alpha,~\beta,~\gamma)$ is there because the associated term
contains three different ways to obtain a singlet, hence the
three couplings (see Appendix~\ref{Delta54}). It should be noted 
that there are in fact two different but equivalent ways to perform 
the product of the first term:

\begin{equation}
(\theta_{1}~\unl{\chi})(\theta_{2}~\unl{\psi})
~~~\text{and}~~~
(\theta_{1}~\theta_{2})(\unl{\chi}~\unl{\psi})
\ .
\end{equation}
Because they are equivalent, there will be no need to
differentiate between them and we shall make no effort in the
future to do so.

For the up-quarks, if we for the moment assumed only one
flavon, say $\theta_1$ with $b=0$, we
see that we respect the $(1,1)$ zero while the $(2,2)$ 
can be filled in. Via FN mechanism we are allowed to have
that $a\approx \delta^2$ so that we can produce the textures
allowed in Eq.~(\ref{eq:oxyukawa1}). A look at our
model will indeed confirm that is what was done.

\item[$(ii)$]{$p=r\neq s$.} The tree-level results should
remain the same. Difference from the results above lie in 
that there are no possible first-order interactions. 

\begin{equation}
\ord(\theta^0):~
\alpha \unl{\chi} ~\unl{\psi} 
+
\beta\chi~\psi
\mapsto
\begin{pmatrix}
0 & \alpha & 0 \\
\alpha & 0 & 0 \\
0 & 0 & \beta \\
\end{pmatrix}
\ .
\end{equation}

The second-order results follows much in the same way as the
case where $p=r=s$; 

\begin{equation}
\ord(\theta^2):~
\left (\alpha,~\beta \right )\theta_1~ \theta_2 ~ \unl{\chi}~\unl{\psi} 
+
\gamma\theta_1~\theta_2~{\chi}~{\psi} 
\mapsto
\begin{pmatrix}
0  & \alpha ad +\beta bc  & 0 \\
\alpha bc +\beta ad & 0   & 0 \\
0  & 0   & \gamma (ad+bc) \\
\end{pmatrix}
\ .
\end{equation}
Once again there is an ambiguity about how to perform the
product of the first term. Direct calculation for all possible
cases shows again that the ambiguity is irrelevant because each
product is equivalent. Notice that there are only two 
couplings, which show that there are only two ways to produce 
singlets for this case.

\end{itemize}

\item{$\mathbf{p\neq r.}$} Now we have the second case of
Eq.~(\ref{eq:matrices}). Before we go on to discuss the two
possible choices, looking at Table~\ref{tb:flavon2}, we find that 

\begin{equation}
\mathbf{2}_p \otimes \mathbf{2}_r 
= 
\mathbf{2}_{s'} \oplus \mathbf{2}_{s''},~~~ p\neq r \neq s'
\neq s'' \  \\
\end{equation}
The above has direct implications at tree-level
since now there is only one result we can have and that is

\begin{equation}
\ord(\theta^0):~
\alpha \chi~\psi
\mapsto
\begin{pmatrix}
0 & 0  & 0 \\
0 & 0 & 0 \\
0 & 0 & \alpha \\
\end{pmatrix}
\ .
\end{equation}
As for the first order, a flavon can only transform as either 
the $2_{s'}$ or the $2_{s''}$. 
The specifics will depend on the representations, but the
results will be in one of four sets of possible combinations
where in each set only one matrix would be chosen:

\begin{equation}
\ord(\theta):~
\alpha \theta_1 ~\unl{\chi}~\unl{\psi}
\mapsto
\begin{array}{ccc}
\begin{pmatrix}
\alpha a & 0  & 0 \\
0 & \alpha b & 0 \\
0 & 0 & 0 \\
\end{pmatrix}
&
 ~~\text{or}~~
&
\begin{pmatrix}
0 & \alpha b  & 0 \\
\alpha a & 0 & 0 \\
0 & 0 & 0 \\
\end{pmatrix}
\\
\\
\begin{pmatrix}
\alpha a & 0  & 0 \\
0 & \alpha b & 0 \\
0 & 0 & 0 \\
\end{pmatrix}
&
 ~~\text{or}~~
&
\begin{pmatrix}
0 & \alpha a  & 0 \\
\alpha b & 0 & 0 \\
0 & 0 & 0 \\
\end{pmatrix}
\end{array}
\ ,
\end{equation}
where we list only two sets for brevity and the other two can
be obtained by interchanging $a$ and $b$. The ``or'' is 
because there are two possible choices for representation of
$\theta_1$, a theme that continues at second order: 

\begin{equation}
\ord(\theta^2):~
\alpha \theta_1 ~\theta_2 ~\unl{\chi}~\unl{\psi}
+
\gamma \theta_1~ \theta_2 ~{\chi}~{\psi}
\mapsto
\begin{array}{l}
\begin{pmatrix}
\alpha ac & 0  & 0 \\
0 & \alpha bd & 0 \\
0 & 0 & \gamma(ad+bc) \\
\end{pmatrix}
~\text{or}~
\begin{pmatrix}
0 & \alpha bd  & 0 \\
\alpha ac & 0 & 0 \\
0 & 0 & \gamma (ad+bc) \\
\end{pmatrix}
\\
\\
\begin{pmatrix}
\alpha ac & 0  & 0 \\
0 & \alpha bd & 0 \\
0 & 0 & \gamma (ad+bc) \\
\end{pmatrix}
~\text{or}~
\begin{pmatrix}
0 & \alpha ac  & 0 \\
\alpha bd & 0 & 0 \\
0 & 0 & \gamma (ad+bc) \\
\end{pmatrix}
\end{array}
\ ,
\end{equation}
where to get the other set of matrices one needs only
interchange the roles of the $ac$ terms with $bd$.

\end{itemize}

The above provides a small glimpse into the workings of two-dimensional 
representations. Although not discussed above one
can tell which entries provide texture zeros by clever choice 
of vevs. With an understanding of the texture structure that
$\Delta(54)$ can produce, we are now ready to discuss our model.

\subsection{Some remarks.} \label{ssect:toymodel}
We had mentioned in the beginning of the section that we would
let the right handed neutrinos transform as $\mathbf{3}_1$ of
our flavor group. The choice is somewhat arbitrary, we could
have easily chosen the representation $\mathbf{\ol{3}}_1$, 
$\mathbf{3}_2$, or $\mathbf{\ol{3}}_2$. Regardless, their Clebsch-Gordan 
(CG) coefficients are similar enough that any choice would do with 
no clear advantage of one over the other. 

As for the choice of two-dimensional representations for the matter 
content, there is some arbitrariness to this too. A look at
Appendix~\ref{Delta54}, focusing on the CG coefficients, will
reveal that all two-dimensional representations under the case
$\mathbf{2}_r\otimes\mathbf{2}_r$ have the same result. The only 
interesting feature occurs on the $\mathbf{2}_p\otimes\mathbf{2}_r$ 
with $p\neq r$ case. One, in terms of model building, could make use 
of the fact that such a product produces two different two-dimensional 
representations. Making it possible to exploit this in a clever fashion, 
but the author has found that using the same two-dimensional 
representation throughout requires less flavons and so a simpler model. 

Finally, now that we have opted to use the same $\mathbf{2}$ for
our model, which one should be used? Looking at
Appendix~\ref{Delta54} shows that taking the product of
$\mathbf{2}_1\times\mathbf{3}_{1}$ produces CG coefficients
that contain powers of $\om=e^{2\pi i/3}$. The same is true for
the cases involving $\mathbf{2}_2$ and $\mathbf{2}_3$ with the
sole exception of $\mathbf{2}_4$. It should be possible to absorb the
$\om$ into coupling constants, thus in effect we have no real
advantage of using one representation over another. However,
for the sake of clarity and simplicity we choose instead to use
$\mathbf{2}_4$ and avoid the issue altogether. 

\section{The $SU(5)\otimes\Delta(54)$ model. }
\cleqn
The model has a supersymmetric background, and we assume that
that we are above unification scale of $SU(5)$ GUT. The matter 
content found in the standard model fits into $SU(5)$ representations as 

\begin{equation}
{X}
\sim
\mathbf{10}
\ ,
~~~
{\Psi}
\sim
\ol{\mathbf{5}}
\ ,
~~~
{\ol{N}}
\sim
\mathbf{1}
\ .
\end{equation}
For reasons discussed in Section~\ref{sect:asflavorgrp} we
chose to have both $\mathbf{\ol{5}}$ and the $\mathbf{10}$ into two and 
one-dimensional representations but kept the heavy neutrinos as 
three-dimensional, i.e. 

\begin{equation}
(X_1,X_2)^{T}\equiv\unl{\chi}\sim \mathbf{2}_4,
~~~
X_3 \equiv \chi \sim \mathbf{1};
~~~
(\Psi_1,\Psi_2)^{T}\equiv\unl{\psi}\sim \mathbf{2}_4,
~~~
\Psi_3 \equiv \psi \sim \mathbf{1}; 
~~~
\ol{N} \sim \mathbf{3}_1
\ .
\end{equation}
Remember that the top quark mass was motivation for using the
singlet and doublet structure for the $\mathbf{10}$. Aside from
these assignments there are other charges that we have given
these fields, namely the $Z^{u}_3\otimes Z^{d}_2\otimes Z_2$
charges. The superscripts indicate that these charges are
primarily given to those fields that contain that right handed particle. 

As we will soon show, the quark and charged lepton sectors are 
populated mainly by three extra fields: 

\begin{equation}
\theta_u \sim \mathbf{2}_4, 
~~~
\theta_d \sim \mathbf{2}_4,
~~~
\sigma \sim \mathbf{1}
\ ,
\end{equation}
The subscripts remind us that these fields are charged under the 
cyclic symmetry ($Z$) with the same letter as its superscript. 

On the other hand, the neutral lepton sector is
primarily populated by just two three-dimensional flavons:

\begin{equation}
\phi \sim \ol{\mathbf{3}}_1, 
~~~
\phi' \sim \ol{\mathbf{3}}_1
\ ,
\end{equation}
once again indicating the appropriate $\Delta(54)$ charge. The 
final ingredients are the Higgs fields which includes both
the five and forty-five dimensional representations of
$SU(5)$. 

We may now present the super-potential, but without all the 
clutter of coupling constants,

\begin{equation}
W_{model}
=
W^{u}+ W^{d}+W^{\nu}_{dirac}+W^{\nu}_{majorana}
\ ,
\end{equation}
where 

\begin{equation}\label{eq:totsups}
\begin{array}{lcl}
W^{u} &\approx&
\chi \chi 
~
H_u
+
(\theta_u ~\unl{\chi} ) \chi 
~
H_{u}
+
\theta_d^2 (\theta_u ~\unl{\chi} ) \chi 
~
H_{u}
+
(\theta_u ~\unl{\chi}) (\theta_u ~\unl{\chi} )
~
H_{u}
+
\theta_d^2(\theta_u ~\unl{\chi}) (\theta_u ~\unl{\chi})
~
H_{u}
\ ,
\\
\\
W^{d}
&\approx&
\chi \psi  
~
H_d
+
( \theta_u ~\unl{\chi}) {\psi}
~
H_{d}
+
\theta_d^2 (\theta_u ~\unl{\chi}) {\psi}
~
H_{d}
+
{\chi}
(\theta_d ~\unl{\psi}) 
~
H_{d}
+
(\theta_u ~\unl{\chi}) (\theta_d ~\unl{\psi}  )
~
H_{d}
+
(\theta_d ~\unl{\psi} ) (\sigma ~\unl{\chi} ~H^{45}_{d} ) \ ,
\\
\\
W^{\nu}_{dirac}
&
\approx
&
\phi \psi \ol{N} 
~
H_u
+
(\phi'~\unl{\psi}) \ol{N}
~
H_u
\ ,
\\
\\
\dfrac{W^{\nu}_{majorana}}{\mathcal M}
&
\approx
&
\phi^2~\ol{N} \ol{N}
+
\phi'^2~ \ol{N} \ol{N}
\ .
\end{array}
\end{equation}
The value $\mathcal{M}$ is the Majorana mass scale that is to
be determined at a later time. We have listed only terms that 
contribute to lowest order in their respective matrix entries. The 
parenthesis have no bearing on how to take products under our flavor 
group, they merely indicate that distinct fields have the correct 
cyclic charges to be neutral under those charges. For a summary of 
the field content and their charges look at Table~\ref{tb:flavon1}.

It should be stated that on Table~\ref{tb:flavon1} we could
have included another cyclic symmetry $Z_2^n$. For this
symmetry the $\ol{N}$ would be odd and so would the $\phi$ and
$\phi'$ flavons. All other fields could in principle remain
neutral. The model however, does not seem to require the extra
symmetry and so we leave this symmetry out of the table.

The next three subsections will contain some of the finer details
of our model. The first two subsections include a look at the vevs of 
the new
fields we have introduced and a detailed look on how each of
the super-potential terms populate their matrices. The last
section presents the final results of our model. These
phenomenological results include the masses for both light and
heavy neutrinos as well as the expected corrections to the
tri-bimaximal matrix. 


\begin{table}[htb]
\caption{Field content and charges of our model with 
         $\om=e^{\frac{2\pi\im}{3}}$. There could be another
	 symmetry $Z_2^n$ but it is found unnecessary. }
\begin{center}
\begin{tabular}{l c c c c c }
\hline
\hline
Matter & $SU(5)$ & $\Delta(54)$ & $Z^u_3$ & $Z^d_{2}$ & $Z_{2}$ \\ 
\hline
\\
$\ol{N}$ & $\mathbf{1}$ & $ \mathbf{3}_1 $ & $ 1 $ & $1$ & $1$ 
\\
$\unl{\psi},~\psi$ &$ \ol{\mathbf 5}$ 
& 
$
\begin{array}{l r}
\mathbf{2}_4,& \mathbf{1} 
\end{array}
$
&
$
\begin{array}{l r}
1, & 1 
\end{array}
$ 
&
$
\begin{array}{l r}
-1, & 1
\end{array}
$
& 
$
\begin{array}{l r}
1, & 1 
\end{array}
$
\\
$\unl{\chi},~\chi$ & $\mathbf{10}$ & 
$
\begin{array}{l r}
\mathbf{2}_4, & \mathbf{1}
\end{array}
$
& 
$
\begin{array}{l r}
\om, & 1
\end{array}
$  
&
$
\begin{array}{l r}
1, & 1 
\end{array}
$
& 
$
\begin{array}{l r}
1 ,& 1
\end{array}
$
\\
\\
Higgs \\ 
\hline
\\
$H_u$, $H_d$ &$ {\mathbf 5}$, $ \ol{\mathbf 5}$& 
$
\begin{array}{l r}
\mathbf{1}, & \mathbf{1}
\end{array}
$ 
& 
$
\begin{array}{l r}
1, & 1
\end{array}
$
&  
$
\begin{array}{l r}
1, & 1
\end{array}
$ 
& 
$
\begin{array}{l r}
1, & 1
\end{array}
$ 
\\
$H^{45}_u$, $H^{45}_d$ & $ {\mathbf{45}}$, $ \ol{\mathbf{45}} $ & 
$
\begin{array}{l r}
\mathbf{1}, & \mathbf{1}
\end{array}
$
& 
$
\begin{array}{l r}
\om^2, & \om
\end{array}
$
& 
$
\begin{array}{l r}
1, & 1
\end{array}
$
& 
$
\begin{array}{l r}
-1, & -1 
\end{array}
$ 
\\
\\
Flavons, $\langle vev \rangle$\\ 
\hline
\\
$\theta_u$,~ 
$\begin{pmatrix}a_1 & 0 \end{pmatrix}^{T}$& $\mathbf{1}$ &
$\mathbf{2}_4$ & $\om^2$ & $1$ & $1$ \\

$\theta_d$,~ 
$\begin{pmatrix}0&a'_2\end{pmatrix}^{T}$ & $\mathbf{1}$ &
$\mathbf{2}_4$ & $1$ & $-1$ & $1$\\

$\phi$,~ 
$\begin{pmatrix}b_1&b_1&0\end{pmatrix}^{T}$ & $\mathbf{1}$ &
$\ol{\mathbf{3}}_1$ & $1$ & $1$ & $1$\\

$\phi'$,~ 
$\begin{pmatrix}b'_1&0&0\end{pmatrix}^{T}$ & $\mathbf{1}$ &
$\ol{\mathbf{3}}_1$ & $1$ & $-1$ & $1$\\

\\
Singlets , $\langle vev \rangle$\\ 
\hline
\\
$\sigma$,~ 
$c$ & $\mathbf{1}$ & $\mathbf{1}$ & $1$ & $1$ & $-1$
\\
\hline
\hline
\end{tabular}
\label{tb:flavon1}
\end{center}
\end{table}

\subsection{Flavon content and vacuum values.}
The vacuum expectation values of the flavon fields go as

\begin{equation}
\langle  \sigma \rangle
\sim c
,
~~~
\langle \theta  \rangle
\sim 
\left ( a_1, ~ a_2 \right )^T
,
~~~
\langle  \phi \rangle
\sim
\left (b_1, ~b_2,~  b_3 \right )^T
\ ,
\end{equation}
where the exact vevs can be found in the table discussed above. 
As said in the introduction, we make use of the
FN mechanism, which means that each flavon vev will be
suppressed by an effective mass scale ($M$) of some gauged
interaction at much higher energies. The suppressed vevs then
are postulated to go as

\begin{equation} \label{eq:vevorder}
\dfrac{c}{M} \approx \delta^{m+1}, ~~~
\dfrac{a_1}{M}\approx \delta^2, ~~~
\dfrac{a'_2}{M}\approx \delta, ~~~
\dfrac{b_1}{M},~ \dfrac{b_1'}{M}\approx \delta^n,~~~
~~~
m \geq 0,~n>0
\end{equation}
where $m$ and $n$ are integers. The value of $m$ can be
determined from the relative size of $v_{5,d}$, the vev of the $H_d$, 
to that $v_{45,d}$ of $H_d^{45}$ by way of

\begin{equation}
\left < \sigma H_d^{45} \right >
\propto
\delta^m v_{45,d}
\approx
v_{5,d}
\ .
\end{equation}
For $v_{45,u}$ we assume that 
$v_{5,u}\geq v_{45,u}\geq v_{45,d}$. The implicit assumption of above is 
that $v_{45,d}\geq v_{5,d}$, otherwise we may lose our perturbative
power by having a singlet with a vev that is greater or equal
to the FN scale. Finally we must mention the relative size of
the $v_{5,u}$ to that of $v_{5,d}$, we expect 

\begin{equation}
cot(\beta)\equiv
\dfrac{v_{5,d}}{v_{5,u}}
\propto
\mathcal{O}
(\delta^{3})
\ ,
\end{equation}
which would satisfy the intra-family relationship $m_b/m_t$.

As for the value of $n$, it may be determined by the size of
the baryon asymmetry our model predicts from leptogenisis
constraints on the lightest of the heavy neutrinos, $M_{1}$
\cite{leptogen}.
Current approximate bounds limit the mass of $M_{1}>10^8~GeV$
and, as we shall see at the end of this section, this limit will 
restrict our possible choices for $n$ such that $n=1,~2,~3$. 

\subsection{Quark Yukawas.}
The purpose of this section is to explore in detail the results
written in Equation~(\ref{eq:totsups}) for the quark sectors. We shall
limit our investigation to demonstrating the origins of all
Yukawa textures and the necessary coupling constants. Each
super-potential contains terms that produce the
leading order contribution to their Yukawa matrix. All other
terms including those which are of $\mathcal{O}(\delta^8)$ and higher
for the up-quarks, and $\mathcal{O}(\delta^5)$ for the
down-quarks, will be neglected. 

The super-potential contributions making the up Yukawa matrix are given by
\begin{eqnarray}\label{eq:yuksuppotent}
W^{u} 
&\approx&
\chi \chi 
~
H_u
+
\alpha (\theta_u ~ \unl{\chi} ) \chi 
~
H_{u}
+
\beta \theta_d^2 (\theta_u ~ \unl{\chi} ) \chi 
~
H_{u}
+
\rho (\theta_u ~\unl{\chi}) (\theta_u ~ \unl{\chi} )
~
H_{u}
+
\\ \nonumber
& &
\gamma \theta_d^2(\theta_u ~ \unl{\chi}) (\theta_u ~ \unl{\chi})
~
H_{u}
\ .
\end{eqnarray}
It should be stated that the $SU(5)$ algebra requires that any 
contribution to the $H_{u}^{45}$ from the $\mathbf{10}$ must be 
anti-symmetric in flavor space. As a result the only anti-symmetric 
combination $(\unl{\chi}~\unl{\chi})_A$ produces a $\mathbf{1}_1$. Since
there are no flavon $\mathbf{1}_1$ singlets there are no
devastating low order contributions, and the only contributions that 
can survive would be corrections to the Yukawa matrices, e.g.
the lowest order correction is: 
$\theta_u^3(\sigma ~\unl{\chi}~\unl{\chi}~H^{45}_{u})$.

In Eq.~(\ref{eq:yuksuppotent}) the Greek letters 
$\alpha,~\beta,~\rho,~\gamma$ are couplings which also aid in 
identifying where each term contributes on the up Yukawa matrix: 

\begin{equation}
{Y}_{5}^{(2/3)}
\approx
\mathcal{O}
\left (
\begin{array}{ccc}
0  &\gamma   \delta^6  &\beta  \delta^4 
\\
\\
\gamma \delta^6 & \rho \delta^4 & \alpha \delta^2 
\\
\\
\beta \delta^4 & \alpha \delta^2 & 1
\end{array}
\right )
\ .
\end{equation}

The down-quark sector is bit more complex for we include both 
contributions due to the regular Higgs $H_d$ and 
the $H_d^{45}$. Both contributions will be added to produce a
single Yukawa matrix, and so below we only include those terms
that are leading in their sum. Primes on Greek letters are for the
couplings that occur on this case, and so the terms we have are

\begin{eqnarray}
W^{d}
&\approx&
\chi \psi  
~
H_d
+
\alpha' ( \theta_u ~\unl{\chi}) {\psi}
~
H_{d}
+
\beta' \theta_d^2 (\theta_u ~\unl{\chi}) {\psi}
~
H_{d}
+
\beta'' {\chi} (\theta_d ~\unl{\psi}) 
~
H_{d}
\\
\nonumber
& & 
+
(\gamma',~\gamma'')(\theta_u ~\unl{\chi}) (\theta_d ~\unl{\psi}  )
~
H_{d}
+
\rho' ~ (\theta_d ~\unl{\psi} ) (\sigma~\unl{\chi}  ~H^{45}_{d})
\ ,
\end{eqnarray}
with $(\gamma',~\gamma'')$ meaning that there are two ways to
produce singlets, each with their own couplings. In terms of
$\delta$ we have

\begin{equation}
{Y}_{5}^{(-1/3)} 
\approx
\mathcal{O}
\left (
\begin{array}{ccc}
0 & \gamma' \delta^3 & \beta' \delta^4 
\\
\\
\gamma'' \delta^3 & 0 & \alpha' \delta^2  \\
\\
\beta'' \delta & 0 & 1
\end{array}
\right )
\ ,
~~~
Y_{45}^{(-1/3)}
\approx
\mathcal{O}
\left (
\begin{array}{ccc}
0 & 0 & 0 \\
\\
0 & \rho' \delta^2 & 0 \\
\\
0 & 0 & 0
\end{array}
\right )
\ .
\end{equation}
Finally with all the above results one can construct the Yukawa matrices 
from the well known results of $SU(5)$ GUT models \cite{georgijarlskog}:

\begin{eqnarray}
Y^{(2/3)}&=&Y^{(2/3)}_5,  \\ \nonumber
Y^{(-1/3)}&=&{Y}_{5}^{(-1/3)} +{Y}_{45}^{(-1/3)},  \\ \nonumber
Y^{(-1)}&=&{Y}_{5}^{(-1/3) T} -3\cdot {Y}_{45}^{(-1/3)} 
\ .
\end{eqnarray}

\subsection{Neutrino masses.}
A similar procedure as outlined in \cite{z7z3} is followed
here. We postulate the addition of two new terms to the super-potential 
of the MSSM:

\begin{equation}
W^{\nu}
=
\mathbf{L} H_u Y^{(0)} \ol{\mathbf N} + \mathcal{M} \ol{\mathbf N}Y_{maj}
\ol{\mathbf N}. 
\end{equation}
The Majorana term also comes with a mass scale $\mathcal{M}$ which we
suppose can come from a higher energy scale. We designed the
model to produce the above with the assumptions that the flavors of
$\mathbf{\ol{N}}$ together form a $\mathbf{3}_1$ under 
our flavor group. To accomplish the task we employed the use of 
two three-dimensional representations $\phi$ and $\phi'$, whose 
details can be found in Table~\ref{tb:flavon1}. 

Our model, Eq.~{\ref{eq:totsups}}, produces the Dirac term 

\begin{eqnarray}
W^{\nu}_{dirac}
&
\approx
&
\phi \psi \ol{N} 
~
H_u
+
(\phi'~\unl{\psi}) \ol{N}
~
H_u
\ ,
\end{eqnarray}
rewritten here for convenience. The resulting Yukawa matrix is

\begin{equation}
Y^{(0)}
\approx
\dfrac{1}{M}
\left (
\begin{array}{ccc}
0 & 0 & b'_1  \\
\\
0 & b'_1 & 0 \\
\\
b_1 & b_1 & 0
\end{array}
\right )
\ .
\end{equation}
For the above there are coupling constants not included because
they are of $\mathcal{O}(1)$ and can be simply absorbed by their
respective vevs. In principle it would be possible to get
tri-bimaximal mixing in the case that 
$\mathcal{O}(b'_1) \neq \mathcal{O}(b_1)$. However if this
is the case and it is carried through to the Majorana matrix then the 
light neutrino matxi $\mathcal{Y}_{\nu}$ would contain entries that 
are sums of various powers in $\delta$. A somewhat simple
calculation will show that this is true. 

In cases like these, it is difficult to diagonalize by 
$\mathcal{U}_{tri-bi}$ since either careful cancellations are needed 
in the various powers in $\delta$ or some explanation for the 
complexity of the coupling constants should be given. To avoid such 
a complication from arising it is found best
to assume that $\mathcal{O}(b'_1)=\mathcal{O}(b_1)$. In fact,
its found that much more elegant results can arise when one
assumes that $b'_1=b_1$ and so this is the assumption we shall make. 

The Majorana contributions terms, found in Eq.~(\ref{eq:totsups}), are 

\begin{eqnarray}
\dfrac{W^{\nu}_{majorana}}{\mathcal M}
&
\approx
&
\phi^2 ~\ol{N} \ol{N}
+
\phi'^2 ~\ol{N} \ol{N}
\ .
\end{eqnarray}
The Majorana matrix is then 

\begin{equation}
\dfrac{Y_{maj}}{\delta^{2n}}
\approx
\begin{pmatrix}
\alpha     & \sigma    & \rho \\
\sigma    & \alpha     & \rho \\
\rho         & \rho          & \beta
\end{pmatrix}
+
\begin{pmatrix}
\alpha' & 0      & 0 \\
0           & 0      & \rho' \\
0           & \rho' & 0
\end{pmatrix}
\ .
\end{equation}
The unprimed Greek letters corresponds to couplings for the
$\phi$ and primed letters for $\phi'$. Do not confuse
these parameters for those written down in the quark sector. 
Just as before they are coupling
constants resulting from the number of ways one can get a
singlet term. Notice that the vevs of the flavons are included,
but found within $\delta^{2n}$ by Eq.~(\ref{eq:vevorder}). The 
best choices for the parameters above seem to be

\begin{equation}
\alpha=\sigma=0,
~~~
\rho=-\beta=\rho'=1,
~~~
|\alpha'|=.100\pm.004
\ .
\end{equation}
The parameter $\alpha'$ can control the value of ratio of the
mass squared differences found in Eq.~(\ref{eq:ratiosdiff}). The 
choice of $|\alpha'|=.1$ produces exactly the ratio of $32$ that 
fits current data.

\subsection{Phenomenological Results.}
We have successfully produced Yukawa matrices with entries of the same
order as we had sought in Eq.~(\ref{eq:oxyukawa1}). We have
even produced a set of matrices for the neutrinos that together 
produce a light neutrino matrix that can be diagonalized by the
tri-bimaximal matrix. Here we take things a step further and
try to reproduce the SM results and find values for neutrino
sector. 

The first step is to reproduce the results of the SM extrapolated to 
the energy scale of $2\times10^{16}~GeV$ \cite{RGstudies}. We have 
seen that for the quark sector, based on our super-potential 
terms, there are ten parameters to be determined. One of these parameters 
if found to be irrelevant and so left equal to one (the $(1,3)$
and $(3,1)$ entries of the up Yukawas). We are then left with
nine that are chosen such that they reproduce masses and the CKM 
matrix which means only seven constraints and so two free parameters.
The last two parameters are chosen such that they at the same
time respect the mass of the down-quark (due to higher order correction) 
and also fit the limits of the experimental results for the solar 
angle of the lepton mixing matrix. Our model has some sensitivity to 
the values of the final parameters which explains the errors we
placed on the predicted angles.

As for the neutrinos we have discussed these free parameters
and because of the constrains imposed by both data and the
tri-bimaximal matrix we have only one parameter (what we called
$\alpha'$ in the neutrino analysis).
\\
\\
{\bf Quark Sector:}
\begin{equation}
Y^{(2/3)}
\approx
\begin{pmatrix}
0 & 1.1 \delta^6 & \delta^4        \\
1.1 \delta^6 &  \delta^4 &   -1.8 \delta^2 \\ 
\delta^4        &  -1.8 \delta^2 &  1      \\
\end{pmatrix}
\ ,
~~~~
m_u
\approx 
v_{5,u}
\begin{pmatrix}
2.7\delta^8 && \\
& 2.3\delta^4 & \\
&&1
\end{pmatrix}
\ ,
\end{equation}
and

\begin{equation}
Y^{(-1/3)}
\approx
\begin{pmatrix}
0        & .5 \delta^3 & .5\delta^4        \\
-.3\delta^3 &  .5\delta^2 & -.6 \delta^2 \\ 
-.5 \delta        & 0 & 1 \\
\end{pmatrix}
\ ,
~~~~
m_d
\approx 
v_{5,d}
\begin{pmatrix}
.6\delta^4 && \\
& .5\delta^2 & \\
&&1
\end{pmatrix}
\ .
\end{equation}
Diagonalization also reproduces a CKM matrix ($U_{ckm}$) 
consistent with data extrapolated to the GUT scale.  
\\
\\
{\bf Lepton Sector:} $SU(5)$ with
$H^{45}_{d}$ guarantees our successful reproduction of the masses

\begin{equation}\label{eq:oxyukawa3}
Y^{(-1)}
\approx
\begin{pmatrix}
0         &  -.3\delta^3 &  -.5\delta \\
.5\delta^3  & -1.5\delta^2  &  0 \\
.5\delta^4         & -.6\delta^2  &  1 \\
\end{pmatrix}
\ ,
~~~~
m_e
\approx 
v_{5,d}
\begin{pmatrix}
.2\delta^4 && \\
& 1.5\delta^2 & \\
&&1
\end{pmatrix}
\ .
\end{equation}

As for the neutrinos we have found that

\begin{equation}
Y^{(0)}
\approx
\delta^{2n}
\begin{pmatrix}
0 & 0 & 1 \\
0 & 1 & 0 \\
1 & 1 & 0 \\
\end{pmatrix}
\ ,
~~~~
Y_{maj}
\approx
\delta^{2n}
\begin{pmatrix}
\alpha' & 0 & 1 \\
0       & 0 & 2 \\
1       & 2 & -1 \\
\end{pmatrix}
,
~~~
|\alpha'|=.100\pm.004
\ ,
\end{equation}
Using the light neutrino approximation and using
$\alpha'=\pm.100$ we obtain 

\begin{equation}\label{eq:oxyukawa4}
\mathcal{Y}_{\nu}
\approx
\dfrac{v_{5,u}^2}{2\mathcal{M}\Delta}
\begin{pmatrix}
0       & \Delta   & \Delta \\
\Delta  & -1       &  1+\Delta \\
\Delta  & 1+\Delta &  -1 \\
\end{pmatrix}
\ ,
~~~~
m_{\nu}
\approx 
\dfrac{v_{5,u}^2}{2\mathcal{M}\Delta }
\begin{pmatrix}
\Delta && \\
& 2\Delta & \\
&& 2+\Delta
\end{pmatrix}
\ ,
\end{equation}
where we remind the reader that $m_{v}$ is the light neutrino 
masses. The value of $\Delta$ is such that $\Delta\approx .222$ for 
$\alpha'=-.100$ and $\Delta\approx-.182$ for $\alpha'=.100$.
We predict that the mass scale $\mathcal {M}$ is

\begin{equation}
\mathcal{M} \approx 3\times10^{15}~GeV
\ ,
\end{equation}
a value that is at one order away from our GUT model scale.
Results that follow are independent on the sign of $\alpha'$.
Both the corrections to the tri-bimaximal matrix and the masses of 
the light neutrinos (normal hierarchy) are predicted to be

\begin{equation}
\begin{array}{cccr}
|\nu_e\rangle &\approx & .825|\nu_1\rangle
+.566|\nu_2\rangle-.127|\nu_3\rangle, & m_{\nu,1}\approx
5\times10^{-3}~eV\\ 
\\
|\nu_{\mu}\rangle &\approx & -.474|\nu_1\rangle
+.532|\nu_2\rangle-.706|\nu_3\rangle, & m_{\nu,2}\approx
1\times10^{-2}~eV\\ 
\\
|\nu_{\tau}\rangle &\approx & -.329|\nu_1\rangle
+.639|\nu_2\rangle+.702|\nu_3\rangle, & m_{\nu,3}\approx
5\times10^{-2}~eV\\ 
\end{array}
\ ,
\end{equation}
where we want to make it clear that 

\begin{equation}
\dfrac{m_{\nu,2}}{m_{\nu,1}} = 2
\ ,
~~~
\dfrac{m_{\nu,3}}{m_{\nu,1}} =10
\ ,
~~~
\text{and}
~~~
\sum_i m_{\nu,i}~ =~ 6.5\times10^{-2}~ eV
\ .
\end{equation}
While we predict that the masses for the heavier neutrinos are

\begin{equation}
M^{\nu}_{heavy}
\approx 
\delta^{2n}
\begin{pmatrix}
9.7\times10^{12}~GeV && \\
& 2.2\times10^{14}~GeV  & \\
&& 3.4\times10^{14}~GeV 
\end{pmatrix}
\ ,
\end{equation}
two masses are nearly degenerate. As mentioned
earlier the value of $n$ could be chosen such that the
masses are consistent with limits posed by leptogenisis
responsible for the baryon asymmetry \cite{leptogen},  

\begin{equation}
M_1\equiv 9.7~\delta^{2n}\times10^{12} ~GeV
>
10^8~GeV
~
\rightarrow
~
n=~1,~2,~3
\ .
\end{equation}

Because corrections for tri-bimaximal matrix are obtain from
diagonalization of the charged lepton Yukawa, care must be
taken so that the angles obtained are well within experimental
limits \cite{neutdata}: 

\begin{equation}
|\theta_{13}|~<~11.4^o,
~~~~
\theta_{\odot} ~\approx ~{34.43^{+1.35}_{-1.22}~}^o,
~~~~
36.8^{o}<-\theta_{atm}<53.2^{o}
\ .
\end{equation}

With the above in mind we predict (and postdict) that 
\begin{equation}
\theta_{13}~\approx~{-7.31^{+0.60}_{-1.75}~}^o,
~~~~
\theta_{\odot} ~\approx ~{34.46^{+1.02}_{-1.52}~}^o,
~~~~
\theta_{atm}~\approx~{-45.15^{+0.04}_{-0.10}~}^o
\ .
\end{equation}
We should mention that the reactor angle ($\theta_{13}$)
is somewhat large. The origin for this is the $(1,3)$ position of 
the charged lepton Yukawa, which leads to a rotation angle 
(from diagonalizing the Yukawa) ``$\theta_{13}$" that 
is comparable to the ``$\theta_{12}$" rotation angle. Now if we
track the phases by following the guidelines given in
\cite{phases}, which provides methods for determining how many free 
phases there are and where in the Yukawas they may be located. We find 
that the $(1,3)$ position for the charged lepton Yukawa could have a 
phase. So the 
reactor angle, being a sum of two comparable angles (as stated earlier) 
with a phase difference between them, could be such that in general
$0^{o}\lesssim -\theta_{13}\lesssim 7.31^{o}$.

\section{A possible modification.}
\cleqn
We present here a modification to our previous model that is based on the
possibility that the flavor singlets of the matter content may be charged 
under the $Z_2$ of our previous model. Table~\ref{tb:flavon2s} contains 
only the changes we expect to make to the model.

\begin{table}[h]
\caption{Changes to the field charges from previous model.}
\begin{center}
\begin{tabular}{l c c c c c }
\hline
\hline
\\
Matter & $SU(5)$ & $\Delta(54)$ & $Z^u_3$ & $Z^d_{2}$ & $Z_{2}$
\\ 
\hline
\\
$\chi$ & $\mathbf{10}$ & 
$ 1 $ & $ 1 $  & $ 1 $ & $ -1 $
\\
$\psi$ &$ \ol{\mathbf 5}$ 
& $ 1 $ & $ 1 $ & $ 1 $ & $ -1 $
\\
\\
Flavons, $\langle vev \rangle$\\ 
\hline
\\
$\phi$,~ 
$\begin{pmatrix}b_1&b_1&0\end{pmatrix}^{T}$ & $\mathbf{1}$ &
$\ol{\mathbf{3}}_1$ & $1$ & $1$ & $-1$\\
\hline
\hline
\end{tabular}
\label{tb:flavon2s}
\end{center}
\end{table}
Notice that one of the $\mathbf 3$ flavons that was previously
neutral is now odd by necessity (unless we change the
neutrino terms) under $Z_2$ charge. As for the super-potential, the 
major changes are the terms that contribute to the $1\times 2$ and 
$2\times 1$ blocks of the Yukawa matrices, no changes are found for 
the neutrino terms.

\begin{equation}
W_{model}
=
W^{u}+ W^{d}+W^{\nu}_{dirac}+W^{\nu}_{majorana}
\ ,
\end{equation}
where 

\begin{equation}\label{eq:totsups1}
\begin{array}{lcl}
W^{u} &\approx&
\chi \chi 
~
H_u
+
(\theta_u ~\unl{\chi} ) (\sigma ~\chi) 
~
H_{u}
+
\theta_d^2 (\theta_u ~\unl{\chi} ) (\sigma ~\chi )
~
H_{u}
+
(\theta_u ~\unl{\chi}) (\theta_u ~\unl{\chi} )
~
H_{u}
+
\theta_d^2(\theta_u ~\unl{\chi}) (\theta_u ~\unl{\chi})
~
H_{u}
\ ,
\\
\\
W^{d}
&\approx&
\chi \psi  
~
H_d
+
\theta_d^2 (\unl{\chi} ~\psi~  H^{45}_{d})
+
(\sigma ~\chi) (\theta_d ~\unl{\psi}) 
~
H_{d}
+
(\theta_u ~\unl{\chi}) (\theta_d ~\unl{\psi}  )
~
H_{d}
+
(\theta_d ~\unl{\psi} ) (\sigma ~\unl{\chi}  ~H^{45}_{d} ) \ ,
\\
\\
W^{\nu}_{dirac}
&
\approx
&
\phi \psi \ol{N} 
~
H_u
+
(\phi'~\unl{\psi}) \ol{N}
~
H_u
\ ,
\\
\\
\dfrac{W^{\nu}_{majorana}}{\mathcal M}
&
\approx
&
\phi^2~\ol{N} \ol{N}
+
\phi'^2~ \ol{N} \ol{N}
\ .
\end{array}
\end{equation}
The vev $\delta^{m+1}$ of the $\sigma$ field still
depends heavily on the relative size of the two Higgs down
vevs. Since we cannot know for sure the value of these, all we can 
do is write down the form of the Yukawa matrix as functions of $m$:

\begin{equation}
{Y}_{5}^{(2/3)}
\approx
\mathcal{O}
\left (
\begin{array}{ccc}
0  &\gamma   \delta^6  &\beta  \delta^{m+5} 
\\
\\
\gamma \delta^6 & \rho \delta^{4} & \alpha \delta^{m+3} 
\\
\\
\beta \delta^{m+5} & \alpha \delta^{m+3} & 1
\end{array}
\right ),
\end{equation}
and 

\begin{equation}
{Y}_{5}^{(-1/3)} 
\approx
\mathcal{O}
\left (
\begin{array}{ccc}
0 & \gamma' \delta^3 & 0
\\
\\
\gamma'' \delta^3 & 0 & 0  \\
\\
\beta'' \delta^{2+m} & 0 & 1
\end{array}
\right )
\ ,
~~~
Y_{45}^{(-1/3)}
\approx
\mathcal{O}
\left (
\begin{array}{ccc}
0 & 0 & 0 \\
\\
0 & \rho' \delta^2 & \alpha'\delta^2 \\
\\
0 & 0 & 0
\end{array}
\right )
\ .
\end{equation}
We have decided to keep the same Greek letters as before
because they still correspond to the same terms of our previous
model with the sole exception of $\alpha'$ which now originates
from the $45$ Higgs. We take as a concrete example the case where $m=1$:
 
\begin{equation}
{Y}^{(2/3)}
\approx
\mathcal{O}
\left (
\begin{array}{ccc}
0  &2.5\delta^6  & \delta^{6} 
\\
\\
2.5\delta^6 & 2.3\delta^{4} & \delta^{4} 
\\
\\
\delta^{6} & \delta^{4} & 1
\end{array}
\right ),
~~~
{Y}^{(-1/3)} 
\approx
\mathcal{O}
\left (
\begin{array}{ccc}
0 & .5 \delta^3 & 0
\\
\\
.6 \delta^3 & .5\delta^2 & 1.3\delta^2  \\
\\
-4\delta^{3} & 0 & 1
\end{array}
\right )
\ .
\end{equation}
Leaving out many of the details and keeping all other results the 
same, the mixing angles for this case of our model become

\begin{equation}
\theta_{13}~\approx~{-1.05^{+2.80}_{-1.16}}^{~o},
~~~~
\theta_{\odot} ~\approx ~{34.48^{+0.52}_{-1.25}}^{~o},
~~~~
\theta_{atm}~\approx~-44.47\pm.01^{~o}
\ .
\end{equation}

\section{Conclusion.}
\cleqn
The goal of this paper was to create a model under $SU(5)$ GUT
that can reproduce all known data with the use of a flavor
group $\Delta(54)$. We began with the SM in
the form of mass hierarchies and one mixing matrix. With these
in mind we found constraints, Eq.~(\ref{eq:oxyukawa1}), on the 
form of the texture structures the quark Yukawa matrices must have. 

A look at the flavor group and the aid of a toy model allowed us to 
see how one can possibly reproduce these texture structures. The 
lepton sector, as far as neutrinos are concerned, was
obtained with a minimalist approach of introducing only the
fewest number of new flavons and fairly simple vev structures.
From these principles we have succeeded in producing a viable
model for neutrinos that can satisfy all constraints provided by
experiments. 

Finally, we provided a possible alternative that would be
viable for a more strict assumptions as to the relationship
between the vevs of the $H_d$ and $H_d^{45}$. The model should
be considered every bit as viable and contains the bonus of
needing less parameters. 

\section{Acknowledgments.}
The author would like to thank Pierre Ramond for his role as a 
mentor, a teacher, and friend. In doing so he has provided many useful
discussions that have contributed to this work. The author would also
like to thank Christoph Luhn for his friendship, patience, advice, and 
his many questions. He has helped many times in both reading 
the author's work and answering questions. This research is partially 
supported by the Department of Energy Grant No. DE-F60297ER41C29.

\clearpage


\section*{Appendix}

\appendix

\section{A comparison of $\Delta(54)$ with $\Delta(27)$.}
\label{Acomp}
\cleqn
There is a great deal of similarities between these two groups,
but $\Delta(27)$ has has been used as a flavor group in a
number of investigations. Likely this has been the case
because, as we shall show here, its structure is not as complex
as that of $\Delta(54)$. The richness in its structure actually
starts with its presentations which shares all the same
features of $\Delta(27)$ but with the addition of two
conjugations and two second-order elements. To see this let's look at the presentation 
for $\Delta(27)$~\cite{D27comp}:

\begin{eqnarray}
\Delta(27)\sim (Z_3 \otimes Z_3) \rtimes Z_3: 
& a^3~=~c^3~=~d^3~=~1, \\ \nonumber
& cd~=~dc, \\ \nonumber
& aca^{-1} ~ =~c^{-1}d^{-1}, ~~ada^{-1}~=~c
\ .
\end{eqnarray}
We clearly see that there are three third-order elements and
as expected two of them commute. Look at Table~\ref{tb:flavon2}
shows that the group includes nine one-dimensional
and two three-dimensional representations. Now the presentation
of $\Delta(54)$~\cite{JACL} is:

\begin{eqnarray}
\Delta(54)\sim (Z_3 \otimes Z_3 ) \rtimes S_3:
& a^3~=~b^2 ~=~(ab)^2 ~=~c^3~=~d^3~=~1, \\ \nonumber
& cd~=~dc, \\ \nonumber
& aca^{-1} ~ =~c^{-1}d^{-1}, ~~ada^{-1}~=~c, \\ \nonumber
& bcb^{-1} ~ = ~ d^{-1}, ~~~~~bdb^{-1}~=~c^{-1}
\ .
\end{eqnarray}
It is clear from the above that $\Delta(54)$ has not only
third-order operators but also second-order ones, which adds to its
complexity. As a result, looking at Appendix~\ref{Delta54}, one sees 
that it has not only one and three-dimensional representations but also 
two-dimensional representations. 

A summary of these facts and a quick description of the
Kronecker products is contained in
Table~\ref{tb:flavon2} found below. 

\begin{table}[htb]
\caption{Summary of some of the differences between
$\Delta(27)$ and $\Delta(54)$. The values 
$r,~ s,~ p,~ t= 1, 2, 3, 4 $}
\begin{center}
\begin{tabular}{c | c}
$\Delta(27)$ & $\Delta(54)$ \\
\hline
\\
nine $\mathbf{1}$- and two $\mathbf{3}$-dimensional reps. 
& two $\mathbf{1}$-, four $\mathbf{2}$-, and four $\mathbf{3}$-dimensional reps.
\\
\\
$
\begin{array}{c}
\mathbf{3} \otimes \mathbf{3} = \ol{\mathbf{3}}\oplus
\ol{\mathbf{3}} \oplus \ol{\mathbf{3}} \\ 
\\
\mathbf{3} \otimes \ol{\mathbf{3}}=\sum^9 \mathbf{1} 
\end{array}
$
& 
$
\begin{array}{c}
\mathbf{2}_p \otimes \mathbf{2}_r 
= 
\mathbf{2}_s \oplus \mathbf{2}_t, p\neq r \neq s \neq t \  \\
\\
\mathbf{2}_r \otimes \mathbf{2}_r = 
(\mathbf{1}\oplus \mathbf{2}_r)_S \oplus \mathbf{1}_{1,A} \\
\\
\mathbf{3}\otimes \mathbf{3} = \ol{\mathbf{3}}\oplus
\ol{\mathbf{3}} \oplus \ol{\mathbf{3}} \\
\\
\mathbf{3}\otimes \ol{\mathbf{3}} = (\mathbf{1}~\text{or}
~\mathbf{1}_1)\oplus \mathbf{2}_1 \oplus \mathbf{2}_2\oplus
\mathbf{2}_3\oplus \mathbf{2}_4
\end{array}
$
\\
\hline
\hline
\end{tabular}
\label{tb:flavon2}
\end{center}
\end{table}


\section{Flavor symmetry $\Delta(54)$.}
\label{Delta54}
\cleqn

The flavor group under consideration is a special case of
$\Delta(6n^2)$,  where $n=3$. A complete study of $\Delta(6n^2)$
can be found in Ref.~\cite{JACL}. From this source we may
obtain the character table, Kronecker products, and the
Clebsch-Gordan coefficients. We list some
results here, specifically the character tables and Kronecker
products.

\subsection{Character table.}
The character table reveals a rich structure behind this 
group. One clearly see that there are one, two and three-dimensional 
representations. Notice that the three-dimensional representations 
are complex, where the conjugates are indicated by a bar. 

\begin{table}[h] 
\begin{center}
\label{tb:characterAa}
\begin{tabular}{c c c c c c c c c c c c} 
\hline
\hline
\multicolumn{11}{c}{Character table of $\Delta(54)$}\\
\hline
\\
$n = 3$ 
& $1C_1$ 
& $1C_1^{(1)}$ 
& $1C_1^{(2)}$ 
& $6C_1$ 
& $6C_2^{(0)}$ 
& $6C_2^{(1)}$ 
& $6C_2^{(2)}$ 
& $9C_3^{(0)}$
& $9C_3^{(1)}$
& $9C_3^{(2)}$\\
\hline
\\
$\mathbf{1}$ & $1$ & $1$ & $1$ & $1$ & $1$ & $1$ & $1$ & $1$ &
$1$ & $1$ \\ 
$\mathbf{1_1}$ & $1$ & $1$ & $1$ & $1$ & $1$ & $1$ & $1$ & $-1$
&
$-1$ & $-1$ \\ 
$\mathbf{2_1}$ & $2$ & $2$ & $2$ & $2$ &  $-1$ & $-1$ & $-1$ &
$0$ & $0$ & $0$\\
$\mathbf{2_2}$ & $2$ & $2$ & $2$ & $-1$& $-1$& $2$& $-1$& $0$&
$0$& $0$\\
$\mathbf{2_3}$ & $2$ & $2$ & $2$ & $-1$ & $-1$& $-1$& $2$& $0$&
$0$& $0$\\
$\mathbf{2_4}$ & $2$ & $2$ & $2$ & $-1$ & $2$& $-1$& $-1$& $0$&
$0$& $0$\\
$\mathbf{3_1}$ & $3$ & $3\om$ & $3\om^2$ & $0$ & $0$ & $0$ &
$0$ & $1$ & $\om^2$ & $\om$ \\
$\mathbf{\bar{3}_1}$ & $3$ & $3\om^2$ & $3\om$ & $0$ & $0$ &
$0$ &
$0$ & $1$ & $\om$ & $\om^2$ \\
$\mathbf{3_2}$ & $3$ & $3\om$ & $3\om^2$ & $0$ & $0$ & $0$ &
$0$ & $-1$ & $-\om^2$ & $-\om$ \\
$\mathbf{\bar{3}_2}$ & $3$ & $3\om^2$ & $3\om$ & $0$ & $0$ &
$0$ &
$0$ & $-1$ & $-\om$ & $-\om^2$ \\
\hline
\hline
\end{tabular}
\caption{ $\om = e^{\frac{2 \pi \im}{3}}$. }
\end{center}
\end{table}

\subsection{Kronecker products.}
In order to build a theory with invariant quantities it's 
necessary to know how products of representations break down 
into irreducible representations. 

\begin{equation}
\begin{array}{ccc}
\begin{array}{ccc}
 {\bf  1_1} \otimes {\bf 1_1} &=& {\bf 1}   \notag \\
 {\bf  1_1} \otimes {\bf 2_1} &=& {\bf 2_1} \notag \\
 {\bf  1_1} \otimes {\bf 2_2} &=& {\bf 2_2} \notag \\
 {\bf  1_1} \otimes {\bf 2_3} &=& {\bf 2_3} \notag \\
 {\bf  1_1} \otimes {\bf 2_4} &=& {\bf 2_4} \notag \\
 {\bf  1_1} \otimes {\bf 3_{1}} &=& {\bf 3_{2}} \notag \\
 {\bf  1_1} \otimes {\bf \bar{3}_{1}} &=& {\bf \bar{3}_{2}}
\notag \\
 {\bf  1_1} \otimes {\bf 3_{2}}&=& {\bf 3_{1}} \notag \\
 {\bf  1_1} \otimes {\bf \bar{3}_{2}}&=& {\bf \bar{3}_{1}}
\notag \\
\end{array}
&
\begin{array}{ccc}
 {\bf  2_1} \otimes {\bf 2_1} &=& 
		    \left ( {\bf 1}+{\bf 2_1} \right )_S
		    +\left ( {\bf 1_1} \right )_A \notag \\ 
 {\bf  2_1} \otimes {\bf 2_2} &=& {\bf 2_3}+{\bf 2_4} \notag \\ 
 {\bf  2_1} \otimes {\bf 2_3} &=& {\bf 2_2}+{\bf 2_4} \notag \\ 
 {\bf  2_1} \otimes {\bf 2_4} &=& {\bf 2_2}+{\bf 2_3} \notag \\ 
 {\bf  2_1} \otimes {\bf 3_{1}}&=& {\bf 3_{1}}+{\bf 3_{2}}
\notag \\
 {\bf  2_1} \otimes {\bf \bar{3}_{1}}&=& {\bf \bar{3}_{1}}+
                    {\bf \bar{3}_{2}} \notag \\
 {\bf  2_1} \otimes {\bf 3_{2}}&=& {\bf 3_{1}}+{\bf 3_{2}}
\notag \\
 {\bf  2_1} \otimes {\bf \bar{3}_{2}}&=& {\bf \bar{3}_{1}}+
                    {\bf \bar{3}_{2}} \notag \\
\end{array}
&
\begin{array}{ccc}
 {\bf  2_2} \otimes {\bf 2_2} &=& 
		    \left (
                    {\bf 1}+{\bf 2_2} \right )_S
		    +\left ( {\bf 1_1}\right )_A \notag \\ 
 {\bf  2_2} \otimes {\bf 2_3} &=& {\bf 2_1}+{\bf 2_4} \notag \\ 
 {\bf  2_2} \otimes {\bf 2_4} &=& {\bf 2_1}+{\bf 2_3} \notag \\ 
 {\bf  2_2} \otimes {\bf 3_{1}} &=& {\bf 3_{1}}+{\bf 3_{2}}
\notag \\
 {\bf  2_2} \otimes {\bf \bar{3}_{1}} &=& {\bf \bar{3}_{1}}+
                    {\bf \bar{3}_{2}} \notag \\
 {\bf  2_2} \otimes {\bf 3_{2}}&=& {\bf 3_{1}}+{\bf 3_{2}}
\notag \\
 {\bf  2_2} \otimes {\bf \bar{3}_{2}}&=& {\bf \bar{3}_{1}}+
                    {\bf \bar{3}_{2}} \notag \\
\end{array}
\end{array}
\end{equation}

\begin{equation}
\begin{array}{cc}
\begin{array}{ccc}

 {\bf  2_3} \otimes {\bf 2_3} &=& 
		    \left (
                    {\bf 1}+{\bf 2_3} \right )_S
		    +\left ({\bf 1_1} \right )_A \notag \\ 
 {\bf  2_3} \otimes {\bf 2_4} &=& {\bf 2_1}+{\bf 2_2} \notag \\ 
 {\bf  2_3} \otimes {\bf 3_{1}} &=& {\bf 3_{1}}+{\bf 3_{2}}
\notag \\
 {\bf  2_3} \otimes {\bf \bar{3}_{1}} &=& {\bf \bar{3}_{1}}+
                    {\bf \bar{3}_{2}} \notag \\
 {\bf  2_3} \otimes {\bf 3_{2}} &=& {\bf 3_{1}}+{\bf 3_{2}}
\notag \\
 {\bf  2_3} \otimes {\bf \bar{3}_{2}} &=& {\bf \bar{3}_{1}}+
                    {\bf \bar{3}_{2}} \notag \\
\end{array}
&
\begin{array}{ccc}
 {\bf  2_4} \otimes {\bf 2_4} &=& 
		    \left (
                    {\bf 1}+{\bf 2_4} \right )_S
		    +
		    \left ( {\bf 1_1} \right )_A\notag \\ 
 {\bf  2_4} \otimes {\bf 3_{1}} &=& {\bf 3_{1}}+{\bf 3_{2}}
\notag \\
 {\bf  2_4} \otimes {\bf \bar{3}_{1}} &=& {\bf \bar{3}_{1}}+
                    {\bf \bar{3}_{2}} \notag \\
 {\bf  2_4} \otimes {\bf 3_{2}} &=& {\bf 3_{1}}+{\bf 3_{2}}
\notag \\
 {\bf  2_4} \otimes {\bf \bar{3}_{2}} &=& {\bf \bar{3}_{1}}+
                    {\bf \bar{3}_{2}} \notag \\
\end{array}
\end{array}
\end{equation}

\begin{eqnarray}
 {\bf  3_1} \otimes {\bf 3_1}&=& 
		    \left (
                    {\bf \bar{3}_{1}}+{\bf \bar{3}_{1}}
		    \right )_S
		    +\left ({\bf \bar{3}_{2}}\right )_A \notag
\\
 {\bf  3_1} \otimes {\bf \bar{3}_1}&=& {\bf 1} +{\bf 2_1}+
                    {\bf 2_2}+{\bf 2_3}+{\bf 2_4} \notag \\
 {\bf  3_1} \otimes {\bf 3_2}&=& {\bf \bar{3}_{1}}+
                    {\bf \bar{3}_{2}}+{\bf \bar{3}_{2}} \notag
\\
 {\bf  3_1} \otimes {\bf \bar{3}_2}&=& {\bf 1_1} +{\bf 2_1}+
                    {\bf 2_2}+{\bf 2_3}+{\bf 2_4} \notag \\
 {\bf  3_2} \otimes {\bf 3_2}&=& 
		    \left (
                    {\bf \bar{3}_{1}}+{\bf \bar{3}_{1}}
		    \right )_S
		    +\left ({\bf \bar{3}_{2}}\right )_A \notag
\\
 {\bf  3_2} \otimes {\bf \bar{3}_2}&=& {\bf 1} +{\bf 2_1}+
                    {\bf 2_2}+{\bf 2_3}+{\bf 2_4} \notag 
 \end{eqnarray}

\subsection{Clebsch-Gordan Coefficients}
We first must define a vector space of each of the irreducible
representations. These will demonstrate how a vector transforms
under the generators a, b, and c of the irreducible
representations. 

\begin{equation}\label{eq:CG1}
\begin{array}{ll}
\begin{array}{ccc}
\mathbf{3}_1~ &:&
\begin{pmatrix} x_1 \\ x_2 \\ x_3 \end{pmatrix}
\mapsto
\begin{pmatrix} x_2 \\ x_3 \\ x_1 \end{pmatrix}_{a},~
\begin{pmatrix} x_3 \\ x_2 \\ x_1 \end{pmatrix}_{b},~
\begin{pmatrix} \om x_1 \\ \om^2 x_2 \\ x_3 \end{pmatrix}_{c}, 
\end{array}
&
\begin{array}{ccc}
\mathbf{\ol{3}}_1~ &:&
\begin{pmatrix} x_1 \\ x_2 \\ x_3 \end{pmatrix}
\mapsto
\begin{pmatrix} x_2 \\ x_3 \\ x_1 \end{pmatrix}_{a},~
\begin{pmatrix} x_3 \\ x_2 \\ x_1 \end{pmatrix}_{b},~
\begin{pmatrix} \om^2 x_1 \\ \om x_2 \\ x_3 \end{pmatrix}_{c},
\end{array}
\\
\\
\begin{array}{ccc}
\mathbf{3}_2~ & : & \begin{pmatrix} x_1 \\ x_2 \\ x_3 \end{pmatrix}
\mapsto
\begin{pmatrix} x_2 \\ x_3 \\ x_1 \end{pmatrix}_a , ~
\begin{pmatrix} -x_3 \\ -x_2 \\ -x_1 \end{pmatrix}_b , ~
\begin{pmatrix} \om x_1 \\ \om^2 x_2 \\ x_3 \end{pmatrix}_c ,
\end{array}
&
\begin{array}{ccc}
\mathbf{\ol{3}}_2~ & : & \begin{pmatrix} x_1 \\ x_2 \\ x_3 \end{pmatrix}
\mapsto
\begin{pmatrix} x_2 \\ x_3 \\ x_1 \end{pmatrix}_a , ~
\begin{pmatrix} -x_3 \\ -x_2 \\ -x_1 \end{pmatrix}_b , ~
\begin{pmatrix} \om^2 x_1 \\ \om x_2 \\ x_3 \end{pmatrix}_c ,
\end{array}
\\
\\
\begin{array}{ccc}
\mathbf{2}_1~ & : & 
\begin{pmatrix} x_1 \\ x_2 \end{pmatrix}
\mapsto
\begin{pmatrix} \om x_1 \\ \om^2 x_2 \end{pmatrix}_a , ~
\begin{pmatrix} x_2 \\ x_1 \end{pmatrix}_b , ~
\begin{pmatrix} x_1 \\ x_2 \end{pmatrix}_c,  
\end{array}
&
\begin{array}{ccc}
\mathbf{2}_2~ & : & 
\begin{pmatrix} x_1 \\ x_2 \end{pmatrix}
\mapsto
\begin{pmatrix} \om x_1 \\ \om^2 x_2 \end{pmatrix}_a , ~
\begin{pmatrix} x_2 \\ x_1 \end{pmatrix}_b , ~
\begin{pmatrix} \om^2 x_1 \\ \om x_2 \end{pmatrix}_c, 
\end{array}
\\
\\
\begin{array}{ccc}
\mathbf{2}_3~ & : & 
\begin{pmatrix} x_1 \\ x_2 \end{pmatrix}
\mapsto
\begin{pmatrix} \om x_1 \\ \om^2 x_2 \end{pmatrix}_a , ~
\begin{pmatrix} x_2 \\ x_1 \end{pmatrix}_b , ~
\begin{pmatrix} \om x_1 \\ \om^2 x_2 \end{pmatrix}_c , 
\end{array}
&
\begin{array}{ccc}
\mathbf{2}_4~ & : & 
\begin{pmatrix} x_1 \\ x_2 \end{pmatrix}
\mapsto
\begin{pmatrix} x_1 \\  x_2 \end{pmatrix}_a , ~
\begin{pmatrix} x_2 \\ x_1 \end{pmatrix}_b , ~
\begin{pmatrix} \om x_1 \\ \om^2 x_2 \end{pmatrix}_c , 
\end{array}
\\
\\
\begin{array}{ccc}
\mathbf{1}_1~ & : & 
\begin{pmatrix} x \end{pmatrix}
~~\mapsto
\begin{pmatrix} x \end{pmatrix}_a ,
~\begin{pmatrix} -x \end{pmatrix}_b ,
~\begin{pmatrix} x \end{pmatrix}_c
\end{array}
\ .
\end{array}
\nonumber
\end{equation}
With the above mappings defined it becomes possible find the
outcomes of taking the product of any two representations. The
list below is not exhaustive, but we include those that are 
important to this paper.  

\cleqn
\begin{itemize}

\item {$x \otimes y $: $\mathbf{1}_1 \otimes \mathbf{1}_1 =
\mathbf{1}$}

\begin{equation}
x \otimes y = xy
\ .
\end{equation}

\item {$x \otimes y $: $\mathbf{1}_1 \otimes \mathbf{2}_r =
\mathbf{2}_r$, ~$r=1,~2,~3,~4$.}

\begin{equation}
x \otimes y = \begin{pmatrix}xy_1\\ -xy_2 \end{pmatrix} \ .
\end{equation}

\item {$x \otimes y $: $\mathbf{2}_r \otimes \mathbf{2}_r =
(\mathbf{1}\oplus\mathbf{2}_r)_S \oplus(\mathbf{1}_1)_A$ }

\begin{equation}
x \otimes y=
\left [
\dfrac{1}{\sqrt{2}}(x_1y_2+x_2y_1) \oplus
\begin{pmatrix} x_2y_2 \\ x_1 y_1\end{pmatrix}
\right ]_S
\oplus
\left [
\dfrac{1}{\sqrt{2}}(x_1y_2-x_2y_1) \oplus
\right]_A
\ .
\end{equation}

\item{$x \otimes y $: $\mathbf{2}_1 \otimes \mathbf{2}_2 =
\mathbf{2}_3 \oplus \mathbf{2}_4$ }

\begin{equation}
x \otimes y=
\begin{pmatrix} x_2y_2 \\ x_1 y_1\end{pmatrix}
\oplus
\begin{pmatrix} x_1y_2 \\ x_2 y_1\end{pmatrix}
\ .
\end{equation}

\item{$x \otimes y $: $\mathbf{2}_1 \otimes \mathbf{2}_3 =
\mathbf{2}_2 \oplus \mathbf{2}_4$ }

\begin{equation}
x \otimes y=
\begin{pmatrix} x_2y_2 \\ x_1 y_1\end{pmatrix}
\oplus
\begin{pmatrix} x_2y_1 \\ x_1 y_2\end{pmatrix}
\ .
\end{equation}

\item{$x \otimes y $: $\mathbf{2}_1 \otimes \mathbf{2}_4 =
\mathbf{2}_2 \oplus \mathbf{2}_3$ }

\begin{equation}
x \otimes y=
\begin{pmatrix} x_1y_2 \\ x_2 y_1\end{pmatrix}
\oplus
\begin{pmatrix} x_1y_1 \\ x_2 y_2\end{pmatrix}
\ .
\end{equation}

\item{$x \otimes y $: $\mathbf{2}_2 \otimes \mathbf{2}_3 =
\mathbf{2}_1 \oplus \mathbf{2}_4$ }

\begin{equation}
x \otimes y=
\begin{pmatrix} x_2y_2 \\ x_1 y_1\end{pmatrix}
\oplus
\begin{pmatrix} x_1y_2 \\ x_2 y_1\end{pmatrix}
\ .
\end{equation}

\item{$x \otimes y $: $\mathbf{2}_2 \otimes
\mathbf{2}_4 = \mathbf{2}_1 \oplus \mathbf{2}_3$ } 

\begin{equation}
x \otimes y=
\begin{pmatrix} x_1y_1 \\ x_2 y_2\end{pmatrix}
\oplus
\begin{pmatrix} x_1y_2 \\ x_2 y_1\end{pmatrix}
\ .
\end{equation}

\item{$x \otimes y $: $\mathbf{2}_3 \otimes \mathbf{2}_4 =
\mathbf{2}_1 \oplus \mathbf{2}_2$ }

\begin{equation}
x \otimes y=
\begin{pmatrix} x_1y_2 \\ x_2 y_1\end{pmatrix}
\oplus
\begin{pmatrix} x_1y_1 \\ x_2 y_2\end{pmatrix}
\ .
\end{equation}

\item{ $x \otimes y $: $\mathbf{2}_1 \otimes \mathbf{3}_1 =
\mathbf{3}_1 \oplus \mathbf{3}_2$ \\ 
$~~~~~~~~~~\mathbf{2}_1 \otimes \mathbf{\ol{3}}_1 =
\mathbf{\ol{3}}_1 \oplus \mathbf{\ol{3}}_2$ 
}

\begin{equation}
x \otimes y=
\dfrac{1}{\sqrt{2}}
\begin{pmatrix} 
x_1y_1 +\om^2 x_2y_1\\ 
\om x_1y_2 +\om x_2y_2\\ 
\om^2 x_1y_3 +x_2y_3
\end{pmatrix}
\oplus
\dfrac{1}{\sqrt{2}}
\begin{pmatrix} 
x_1y_1 -\om^2 x_2y_1\\ 
\om x_1y_2 -\om x_2y_2\\ 
\om^2 x_1y_3 -x_2y_3
\end{pmatrix}
\ .
\end{equation}

\item{ $x \otimes y $: $\mathbf{2}_4 \otimes \mathbf{3}_1 =
\mathbf{3}_1 \oplus \mathbf{3}_2$ }

\begin{equation}
x \otimes y=
\dfrac{1}{\sqrt{2}}
\begin{pmatrix} 
x_1y_3 + x_2y_2\\ 
x_1y_1 + x_2y_3\\ 
x_1y_2 + x_2y_1
\end{pmatrix}
\oplus
\dfrac{1}{\sqrt{2}}
\begin{pmatrix} 
x_1y_3 - x_2y_2\\ 
x_1y_1 - x_2y_3\\ 
x_1y_2 - x_2y_1
\end{pmatrix}
\ .
\end{equation}

\item{ $x \otimes y $: $\mathbf{2}_4 \otimes \mathbf{\ol{3}}_1 =
\mathbf{\ol{3}}_1 \oplus \mathbf{\ol{3}}_2$ }

\begin{equation}
x \otimes y=
\dfrac{1}{\sqrt{2}}
\begin{pmatrix} 
x_1y_2 + x_2y_3\\ 
x_1y_3 + x_2y_1\\ 
x_1y_1 + x_2y_2
\end{pmatrix}
\oplus
\dfrac{1}{\sqrt{2}}
\begin{pmatrix} 
x_1y_2 - x_2y_3\\ 
x_1y_3 - x_2y_1\\ 
x_1y_1 - x_2y_2
\end{pmatrix}
\ .
\end{equation}

\item{ $x \otimes y $: $\mathbf{3}_1 \otimes \mathbf{3}_1 =
(\mathbf{\ol{3}}_1 \oplus \mathbf{\ol{3}}_1)_S \oplus 
(\mathbf{\ol{3}}_2)_A$ }

\begin{equation}
x \otimes y =
\left [
\begin{pmatrix} 
x_1y_1 \\
x_2y_2 \\
x_3y_3
\end{pmatrix}
\oplus
\dfrac{1}{\sqrt{2}}
\begin{pmatrix} 
x_2y_3 + x_3y_2\\ 
x_3y_1 + x_1y_3\\ 
x_1y_2 + x_2y_1
\end{pmatrix}
\right ]_S
\oplus
\left [
\dfrac{1}{\sqrt{2}}
\begin{pmatrix} 
x_2y_3 - x_3y_2\\ 
x_3y_1 - x_1y_3\\ 
x_1y_2 - x_2y_1
\end{pmatrix}
\right ]_A 
\ .
\end{equation}

\item{ $x \otimes y $: $\mathbf{3}_1 \otimes \mathbf{\ol{3}}_1 =
\mathbf{1}\oplus \mathbf{2}_1\oplus \mathbf{2}_2 \oplus \mathbf{2}_3 \oplus
\mathbf{2}_4 $}

\begin{eqnarray}
x \otimes y & = & 
\dfrac{1}{\sqrt{3}} \left (x_1y_1+ x_2y_2+x_3y_3 \right )
~ \oplus ~
\dfrac{1}{\sqrt{3}} 
\begin{pmatrix} 
x_1y_1 +\om^2 x_2y_2+\om x_3y_3 \\
\om x_1y_1 + \om^2 x_2y_2 +x_3 y_3 
\end{pmatrix}\\  \nonumber
& & \oplus ~
\dfrac{1}{\sqrt{3}}
\begin{pmatrix} 
x_1y_2 + \om^2 x_2y_3 + \om x_3y_1 \\ 
x_3y_2 + \om^2 x_2y_1 + \om x_1y_3
\end{pmatrix}
~\oplus ~
\dfrac{1}{\sqrt{3}}
\begin{pmatrix} 
x_2y_1 + \om^2 x_3y_2 + \om x_1y_3 \\ 
x_2y_3 + \om^2 x_1y_2 + \om x_3y_1 
\end{pmatrix} \\ \nonumber
&  &\oplus ~
\dfrac{1}{\sqrt{3}}
\begin{pmatrix} 
x_3y_2 + x_2y_1 + x_1y_3 \\ 
x_2y_3 + x_1y_2 + x_3y_1 
\end{pmatrix}
\ .
\end{eqnarray}

\item{ $x \otimes y $: $\mathbf{3}_1 \otimes \mathbf{3}_2 =
\mathbf{\ol{3}}_2 \oplus \mathbf{\ol{3}}_2 \oplus \mathbf{\ol{3}}_1$ }

\begin{equation}
x \otimes y =
\begin{pmatrix} 
x_1y_1 \\
x_2y_2 \\
x_3y_3
\end{pmatrix}
\oplus
\dfrac{1}{\sqrt{2}}
\begin{pmatrix} 
x_2y_3 + x_3y_2\\ 
x_3y_1 + x_1y_3\\ 
x_1y_2 + x_2y_1
\end{pmatrix}
\oplus
\dfrac{1}{\sqrt{2}}
\begin{pmatrix} 
x_2y_3 - x_3y_2\\ 
x_3y_1 - x_1y_3\\ 
x_1y_2 - x_2y_1
\end{pmatrix}
\ .
\end{equation}

\item{ $x \otimes y $: $\mathbf{3}_1 \otimes \mathbf{\ol{3}}_2 =
\mathbf{1}_1\oplus \mathbf{2}_1\oplus \mathbf{2}_2 \oplus 
\mathbf{2}_3 \oplus \mathbf{2}_4 $}

\begin{eqnarray}
x \otimes y & = & 
\dfrac{1}{\sqrt{3}} \left (x_1y_1+ x_2y_2+x_3y_3 \right )
~\oplus~
\dfrac{1}{\sqrt{3}} 
\begin{pmatrix} 
x_1y_1 +\om^2 x_2y_2+\om x_3y_3 \\
-\om x_1y_1 - \om^2 x_2y_2 -x_3 y_3 
\end{pmatrix}\\  \nonumber
& & \oplus~
\dfrac{1}{\sqrt{3}}
\begin{pmatrix} 
x_1y_2 + \om^2 x_2y_3 + \om x_3y_1 \\ 
-x_3y_2 - \om^2 x_2y_1 - \om x_1y_3
\end{pmatrix}
~\oplus~
\dfrac{1}{\sqrt{3}}
\begin{pmatrix} 
-x_2y_1 - \om^2 x_3y_2 - \om x_1y_3 \\ 
x_2y_3 + \om^2 x_1y_2 + \om x_3y_1 
\end{pmatrix} \\ \nonumber
&  &\oplus~
\dfrac{1}{\sqrt{3}}
\begin{pmatrix} 
-x_3y_2 - x_2y_1 - x_1y_3 \\ 
x_2y_3 + x_1y_2 + x_3y_1 
\end{pmatrix}
\ .
\end{eqnarray}

\item{ $x \otimes y $: $\mathbf{\ol{3}}_1 \otimes
\mathbf{\ol{3}}_1 = (\mathbf{3}_1 \oplus 
\mathbf{3}_1)_S \oplus (\mathbf{3}_2)_A$ }

\begin{equation}
x \otimes y =
\left [
\begin{pmatrix} 
x_1y_1 \\
x_2y_2 \\
x_3y_3
\end{pmatrix}
\oplus
\dfrac{1}{\sqrt{2}}
\begin{pmatrix} 
x_2y_3 + x_3y_2\\ 
x_3y_1 + x_1y_3\\ 
x_1y_2 + x_2y_1
\end{pmatrix}
\right ]_S
\oplus
\left [
\dfrac{1}{\sqrt{2}}
\begin{pmatrix} 
x_2y_3 - x_3y_2\\ 
x_3y_1 - x_1y_3\\ 
x_1y_2 - x_2y_1
\end{pmatrix}
\right ]_A \ .
\end{equation}

\end{itemize}




\end{document}